%% file: lectures.tex
\documentclass[12pt]{article}

\usepackage{amsmath,amssymb,graphicx}
\numberwithin{equation}{section}
\usepackage{cite}

\newcommand{\beq}{\begin{eqnarray}}
\newcommand{\eeq}{\end{eqnarray}} 
\newcommand{\bra}[1]{\left\langle #1\right|}
\newcommand{\ket}[1]{\left|#1\right\rangle}

\newcommand{\vev}[1]{\langle #1\rangle}
\newcommand{\Order}[1]{\ensuremath{\mathcal{O}\left(#1\right)}}

\def\Psibar{\overline \Psi }
\def\psibar{\bar\psi}
\def\chibar{\bar\chi}

\def\alphadot{{\dot\alpha}}
\def\betadot{{\dot\beta}}
\def\thetabar{{\bar\theta}}
\def\sigmabar{{\bar\sigma}}
\def\fprime{f^\prime}

\def\Rsb{{R\!\!\!\!\slash}}

\def\Mpl{M_\mathrm{Pl}}
\def\LQCD{\Lambda_\mathrm{QCD}}
\def\Lagr{\mathcal{L}}
\def\pd{\partial}
\def\kahler{K\"ahler }
\def\oraf{O'Raifeartaigh }
\def\Wg{\mathcal{W}}

\def\Tr{\mathrm{Tr}}
\def\Str{\mathrm{Str}}
\def\hc{\mathrm{h.c.}}
\def\goldstino{{\tilde G}}
\def\TeV{\mathrm{TeV}}
\def\GeV{\mathrm{GeV}}
\def\Pf{\ensuremath{\mathrm{Pf}}}
\def\unit{\relax{\rm 1\kern-.26em I}}
\newcommand{\drawsquare}[2]{\hbox{%
\rule{#2pt}{#1pt}\hskip-#2pt
\rule{#1pt}{#2pt}\hskip-#1pt
\rule[#1pt]{#1pt}{#2pt}}\rule[#1pt]{#2pt}{#2pt}\hskip-#2pt
\rule{#2pt}{#1pt}}
\newcommand{\fund}{\drawsquare{6.5}{0.4}}
\newcommand{\afund}{\overline{\fund}}

\begin{document}
\begin{titlepage}
\begin{flushright}
\end{flushright}    

\vskip.5cm
\begin{center}
{\huge \bf TASI 2008 Lectures:\\ Introduction to Supersymmetry\\ and Supersymmetry Breaking \\}

\vskip.1cm
\end{center}
\vskip0.2cm 

\begin{center}
{\bf 
Yuri Shirman}          
\end{center}                 
\vskip 8pt

\begin{center}
 {\it Department of Physics and Astronomy\\ University of California, Irvine,
CA                                                                  
92697.}                                                

\vspace*{0.3cm}                                           
{\tt   yshirman@uci.edu}
\end{center}                                                                 

\vglue 0.3truecm

\begin{abstract}
\vskip 3pt \noindent These lectures, presented at TASI 08 school, provide an introduction to supersymmetry and supersymmetry breaking. We present basic formalism of supersymmetry, supersymmetric non-renormalization theorems, and summarize non-perturbative dynamics of supersymmetric QCD. We then turn to discussion of tree level, non-perturbative, and metastable supersymmetry breaking. We introduce Minimal Supersymmetric Standard Model and discuss soft parameters in the Lagrangian. Finally we discuss several mechanisms for communicating the supersymmetry breaking between the hidden and visible sectors.
\end{abstract}

\end{titlepage}

\input{supersymmetry}

%
%
%
%
%

\end{document}

%% file: supersymmetry.tex
\tableofcontents
\section{Introduction}
\subsection{Motivation}
In this series of lectures we will consider introductory topics in the study of supersymmetry (SUSY) and supersymmetry breaking. There are many motivations which make SUSY a worthwhile subject of research. Since the topic of this TASI school is LHC, we will concentrate only on the motivation most closely related to physics at the TeV scale that will be probed by LHC experiments --- the gauge hierarchy problem.

The Standard Model of particle physics is a consistent quantum field theory that may be valid up to energies as high as $\Mpl$. On the other hand, it is also characterized by some intrinsic energy scales such as $\LQCD$ and the scale of electroweak 
symmetry breaking (EWSB), $m_Z\sim 100\, \GeV$, both of which are much smaller than $\Mpl$. There is nothing disturbing about the smallness of the ratio $\LQCD/\Mpl$. Indeed, $\LQCD$ is generated by dimensional transmutation from the dimensionless parameter $g^2$, QCD coupling constant. The value of $\LQCD$ is exponentially sensitive to the value of the gauge coupling at the cutoff scale $M$, $\LQCD\sim M \exp(-8\pi^2/b_og^2(M))$, where $b_o$ is a one loop $\beta$-function coefficient. A small $\Lambda_{QCD}$ can be obtained with $\Order{1}$ coupling at the cutoff scale, even when the cutoff is taken to be $\Mpl$.  On the other hand, $Z$ mass is determined by the Higgs vacuum expectation value (vev). This poses a conceptual problem since  quantum corrections generically make mass parameters in the scalar field Lagrangian as large as the cutoff scale of the theory. Let us illustrate this with a simple example.
Consider a Yukawa model with a massless scalar field:
\begin{equation}
 \Lagr=\frac{1}{2}(\pd_\mu\varphi)^2+i\psibar\gamma^\mu\pd_\mu\psi-\frac{\lambda}{4!}\varphi^4-y\varphi\psibar\psi\,.
\end{equation}

\begin{figure}[t]
\begin{center}
 \includegraphics[width=5in]{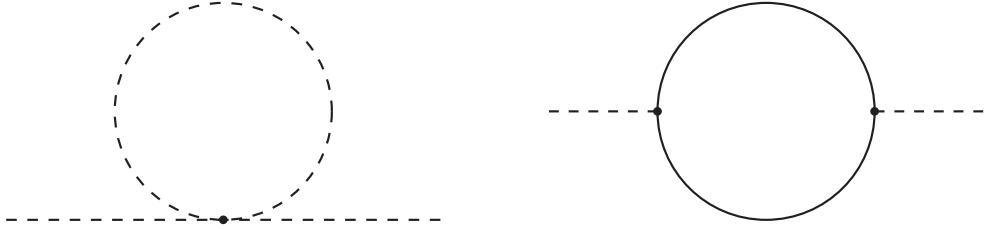}
\end{center}
\caption{One loop diagrams contributing to scalar mass squared in Yukawa theory.}
\label{figscalarm2}
\end{figure}

We can easily calculate renormalization of the scalar mass squared. At one loop order there are two contributions arising from the scalar and fermion loops shown in Fig. \ref{figscalarm2}. Both contributions are individually quadratically divergent:
\begin{equation}
\begin{aligned}
&-i \delta m^2|_\text{scalar loop}\sim -i \frac{\lambda}{32\pi^2}M^2\,,\\
&-i \delta m^2|_\text{fermion loop}\sim i \frac{4y^2}{16\pi^2}M^2\,.
\end{aligned}
\end{equation}
Generically (for $\lambda\ne y^2/8$) this leads to a quadratically divergent contribution to scalar mass squared, just one loop below the cutoff of the theory
\beq
m^2\sim \frac{\lambda/2-4y^2}{16\pi^2}M^2\,.
\eeq
If our goal is to construct a low energy theory with a light scalar particle, $m\ll M$, we need to make sure this correction cancels the bare mass in the classical Lagrangian to a very high precision. This is equivalent to fine-tuning the Lagrangian parameters $m_0$, $\lambda$, and $y$. Even if we do so at one loop, two loop corrections will be quadratically divergent again. In general, adjusting coupling constants so that the leading contribution appears only at $n$-loop order simply suppresses the mass squared correction by a factor of the order  $(1/4\pi)^n$ relative to the UV scale $M$ (assuming order one coupling constants). 
The need for such a cancellation implies that the low energy physics is sensitive to arbitrarily high energy scales. The presence of additional heavy particles with masses of order $M$ can modify $\lambda$ and $y$ and affect cancellations of quadratic divergencies that low energy theorist worked so hard to arrange. As a result the mass of our light scalar $\varphi$ will sensitively depend on the physics at arbitrarily high energy scales.
In the Standard Model a similar problem, usually referred to as a gauge hierarchy problem, requires an explanation of a hierarchy of some 17 orders of magnitude between the scale of electroweak symmetry breaking  and the Planck scale.

It is useful to compare mass renormalization of a scalar field to that of a fermion.  Theories of massless fermions possess a chiral symmetry which forbids mass terms. Thus mass terms can not be generated radiatively unless the symmetry is broken -- for example by tree level masses. In such a case radiative corrections to fermion masses are proportional to their tree level values and can be at most logarithmically divergent. While the one loop contribution to the fermion mass is enhanced by large logs, it can remain small even when the cutoff scale is as large as $\Mpl$. This mechanism does not explain the origin of a hierarchy between fermion masses and UV scales in the theory, but once introduced into the theory the hierarchy is not destabilized by radiative effects. Thus mass hierarchies in the fermion sector are at least {\em technically natural}. In fact, in some cases fermion masses arise dynamically. For example, proton and neutron masses are largely determined by strong QCD dynamics and naturally are of the order $\Lambda_{QCD}$ which, in turn, can easily be small compared to the Planck scale.
In this case, not only is the fermion mass stable against small changes in parameters of the theory, it is also naturally small --- QCD dynamics explains the {\em origin} of the hierarchy between baryon masses and Planck scale. 

We would like to find similar explanations for the origin and stability of the EWSB scale. In particular, we would like to find an extension of the Standard Model with new physics at the TeV. The presence of new fields and interactions would provide a cutoff for the Standard Model calculation of quantum corrections to the Higgs mass. If the scale of new physics is stable against radiative corrections, the technical naturalness problem would be resolved. 
We would further like to find a theory which also explains the origin of the hierarchy between TeV and Planck scales. Whether this situation is realized in nature or the specific value of electroweak scale is just an ``accident'' will eventually be answered by experimental data.

Over the years significant effort was invested into the investigation of possible
solutions of the naturalness problem.
Supersymmetry, the symmetry relating particles with different spins, 
emerged as one of the leading candidates for such a solution.
If the Standard Model were supersymmetric, one could easily explain the stability of gauge hierarchy. Indeed, in a supersymmetric model the Higgs boson would be related by symmetry to a spin $\textstyle{1/2}$ particle (particles related by SUSY are called superpartners, and the superpartner of the Higgs boson is referred to as higgsino), and this symmetry would guarantee that the masses of bosonic and fermionic partners are equal. On the other hand, corrections to the  fermion mass are at most logarithmically divergent --- thus in a supersymmetric theory scalar particles may be naturally light. Of course, we have not yet observed even a single elementary scalar particle, let alone a particle with the same mass as the mass of any known fermion. Therefore at low energies supersymmetry may only be an approximate symmetry. The low energy effective Lagrangian must contain SUSY breaking terms. On the other hand, adding an arbitrary SUSY breaking term to the Lagrangian  would be dangerous -- one must take care not to reintroduce quadratic divergencies. Therefore, while supersymmetry can not be a symmetry of the ground state, it must remain a symmetry of the Lagrangian; it must be broken spontaneously rather than explicitly.

As we will see later, supersymmetry is extremely difficult to break. It can be broken either at tree level or, in some theories, non-perturbatively. In the latter case the scale of dynamical effects leading to supersymmetry breaking can naturally be low providing  an explanation not only for the stability but also for the origin of gauge hierarchy.

This series of lectures is intended as a first introduction to supersymmetry and supersymmetry breaking. We begin by briefly reviewing Weyl fermions and introducing SUSY algebra. In section \ref{sec:SUSY} we  construct supersymmetric Lagrangians step by step starting with Wess-Zumino model and progressing to non-abelian SUSY gauge theories. In section \ref{sec:NRtheorem} we discuss non-renormalization theorems which provide powerful tools in theoretical studies of SUSY. We also review non-perturbative dynamics of non-abelian SUSY gauge theories. In section \ref{sec:DSB} we discuss spontaneous and dynamical SUSY breaking. We then introduce Minimal Supersymmetric Standard Model in section \ref{sec:MSSM}. Finally, in section \ref{sec:mediation} we discuss interactions between MSSM sector and so called hidden sector where SUSY must be broken.

\subsection{Weyl fermions}
Let us briefly review the description of theories with fermions in a language of two component Weyl spinors. Consider the theory of a free Dirac fermion
\beq
\Lagr=i\Psibar \pd_\mu \gamma^\mu\psi -m \Psibar \Psi\,.
\eeq

It is convenient to work in a chiral basis where $\gamma$-matrices take the form
\beq
\gamma^\mu=\left(\begin{array}{cc}
0&\sigma^\mu\\\sigmabar^\mu &0
\end{array}\right), ~~~
                    \gamma^5= \left(\begin{array}{cc}
                      -1_2&0\\
                        0&-1_2\\
                     \end{array}\right),~~~
~~~\sigma^\mu=(1_2,\sigma^i), ~~~\sigmabar^\mu=(1_2,-\sigma^i)\,,
\eeq
and $\sigma^i$ are Pauli matrices
\begin{equation}
 \sigma^1=\left(\begin{array}{cc}
 0&1\\
1&0\\
\end{array}
\right),~~~\sigma^2=\left(\begin{array}{cc}
 0&-i\\
i&0\\
\end{array}
\right),~~~
\sigma^3=\left(\begin{array}{cc}
 1&0\\
0&-1\\
\end{array}
\right)\,.
\end{equation}

We can be explicit about left and right handed components of $\Psi$ by  writing
\begin{equation}
 \Psi=\left(\begin{array}{c} \eta^\alpha\\ \chibar^*_\alphadot \end{array}\right)\,.
\end{equation}
Here $\eta^\alpha$ and $\chibar^*_\alphadot$ are left and right handed two-component fermions and $\alpha,\alphadot=1,2$. If $\Psi$ has well defined transformations under some local or global symmetry, charges of $\eta$ and $\chibar^*$ under this symmetry are identical. On the other hand, $\eta$ and $\chibar$ are both left-handed but transform in conjugate representations under all the symmetries\footnote{In these lectures bar above the quantum field denotes a conjugate representation under (relevant) symmetry rather than complex conjugation. For example, $\chi$ and $\bar \chi$ represent two different fields with opposite charges, while $\chi^*$ and $\bar\chi^*$ are antiparticles of $\chi$ and $\bar \chi$ respectively.}.

It is convenient to lower and raise spinor indices with $\epsilon$-tensors $\epsilon^{\alpha\beta}$ and $\epsilon_{\alphadot\betadot}$. Lorentz scalars can be written as
\begin{equation}
\begin{array}{lll}
&\eta^\alpha\eta_\alpha=\epsilon^{\alpha\beta}\eta_\alpha\eta_\beta,
~~~&\eta^*_\alphadot\eta^{*\alphadot}=\epsilon_{\alphadot\betadot}\eta^{*\alphadot}\eta^{*\betadot},~~~\\ &\eta\chibar=\chibar\eta=
\epsilon^{\alpha\beta}\eta_\alpha\chibar_\beta,~~~ &\eta^*\chibar^*=\chibar^*\eta^*=
\epsilon_{\alphadot\betadot}\eta^\alphadot\chibar^\betadot\,.
\end{array}
\end{equation}
Left-handed and right-handed spinors can be combined into Lorentz vectors as 
\begin{equation}
 \eta^*_\alphadot\sigmabar^{\mu\alphadot\alpha}\eta_\alpha=-\eta^\alpha\sigma^\mu_{\alpha\alphadot}\eta^{*\alphadot}\,.
\end{equation}
We can now write the Lagrangian in terms of left-handed Weyl fermions $\eta$ and $\chibar$
\begin{equation}
 \Lagr=i\eta^*\pd_\mu\sigmabar^\mu \eta+i\chibar^*\pd_\mu \sigmabar^\mu\chibar-m\chibar\eta-m\chibar^*\eta^*,
\end{equation}
where we have integrated by parts to obtain identical kinetic terms for $\eta$ and $\chibar$.
Clearly we can write down theories with both gauge and Yukawa interactions in terms of Weyl fermions. 
In fact Weyl fermions represent the most natural language for the description of chiral theories.

\subsection{A first look at supersymmetry}
Our goal is to construct a quantum field theory with the symmetry relating fermions and bosons. This symmetry is referred to as {\em supersymmetry}. The generator of the symmetry must relate two types of particles:
\beq
Q\ket{\mbox{fermion}}=\ket{\mbox{boson}},~~~Q\ket{\mbox{boson}}=\ket{\mbox{fermion}}\,.
\eeq
It follows that $Q$ must be a spinor. Furthermore, in 4-dimensional spacetime the minimal spinor is a Weyl spinor and therefore the minimal supersymmetry has 4 supercharges. 

In fact, under quite general assumptions there exist a unique non-trivial extension of the Poincare symmetry \cite{Coleman:1967ad} which includes spinorial generators. This extension forms graded-Lie algebra \cite{Golfand:1971iw} defined by the usual commutation relations of the Poincare symmetry together with the new anti-commutation relations:
\begin{equation}
\label{eq:susyalgebra}
\begin{aligned}
&\{Q_\alpha, \bar Q_{\dot\beta}\}=2\sigma_{\alpha\dot\beta}^\mu P_\mu \,,\\
&\{Q_\alpha, Q_\beta\}=\{\bar Q_{\dot\beta},\bar Q_{\dot\beta}\}=0\,,\\
&[Q_\alpha, P_\mu]=[\bar Q_{\dot\beta}, P_\mu]=0\,,
\end{aligned}
\end{equation}
where $P$ is the translation generator.

Even before we begin our study of supersymmetric Lagrangians
the algebra (\ref{eq:susyalgebra}) leads to important consequences.
Taking the trace on both sides of the first equation in (\ref{eq:susyalgebra}) we find
\beq
Q_1\bar Q_{\dot 1}+\bar Q_{\dot 1} Q_1+Q_2\bar Q_{\dot 2}+\bar Q_{\dot 2} Q_2=4 P^0 \,.
\eeq

Any state in the theory that is invariant under a symmetry is annihilated by symmetry generators. In particular, if the ground state $\ket{0}$ is supersymmetric, it is annihilated by SUSY generators. In general for an energy eigenstate we have
\beq
\bra{E}Q_1\bar Q_{\dot 1}+\bar Q_{\dot 1} Q_1+Q_2\bar Q_{\dot 2}+\bar Q_{\dot 2} Q_2\ket{E}=\bra{E}4 P^0\ket{E}=4E\,. 
\eeq
We, therefore, conclude that the energy of a supersymmetric ground state must be zero. On the other hand, if supersymmetry is spontaneously broken, the vacuum energy is positive definite.

Our last statement is strictly correct only in the limit $M_{Pl}\rightarrow \infty$. In a theory with dynamical gravity supersymmetry must be a local symmetry. In this case energy of supersymmetric ground states ({\em i.e.} cosmological constant) is non-positive but not necessarily zero. This means that  while ground states with a positive energy can not be supersymmetric, non-supersymmetric ground states do not necessarily have positive energy.

\section{Constructing supersymmetric Lagrangians}
\label{sec:SUSY}
\subsection{Wess-Zumino Model}
\label{sec:WZModel}
Our first goal is to construct the simplest possible supersymmetric field theory. Such a theory must contain at least one Weyl fermion, $\eta$, a minimal spinor in 4 dimensions. As its superpartner we need either a complex field, $\varphi$, or a vector field, $A_\mu$. Let us use the complex scalar for our first example of supersymmetry. 
An on-shell 
Weyl fermion contains two degrees of freedom. Therefore, its spin 0  superpartner  must have two degrees of freedom, {\em i.e.} it must be a complex scalar. 
However, an off-shell Weyl fermion has 4 degrees of freedom while a complex scalar still has two degrees of freedom. If we want to maintain explicit supersymmetry while performing calculations involving off-shell particles, number of bosonic and fermionic degrees of freedom must match both on-shell and off-shell. Thus we need to add an additional, {\em auxiliary}, complex bosonic field, $F$, which has 2 degrees of freedom off-shell and no on-shell degrees of freedom. This can be achieved by introducing a field with purely algebraic equations of motion, that is a field without kinetic terms.
Let's try the simplest non-interacting Lagrangian with all the required degrees of freedom (this theory \cite{Wess:1974jb}
is known as free Wess-Zumino model):
\beq
\label{eq:free}
\Lagr=\int d^4x |\pd_\mu \varphi|^2 + i\eta^*\pd_\mu \sigmabar^\mu \eta + |F|^2\,.
\eeq
Note that the $F$ equation of motion is indeed algebraic and sets $F=0$ in the ground state of the theory.
We now specify how supersymmetry transformations act on the fields:
\beq
\label{eq:SUSYtransforms}
\delta \varphi =\epsilon^\alpha \eta_\alpha\,,&~~~&\delta \varphi^*=\epsilon^*_\alphadot \eta^{*\alphadot}\,,\nonumber\\
\delta\eta_\alpha=-i \sigma^\mu_{\alpha\alphadot}\epsilon^{*\alphadot}\pd_\mu \varphi+\epsilon_\alpha F\,,&~~~&\delta\eta_\alphadot^*=i\epsilon^\alpha\sigma^\mu_{\alpha\alphadot}\pd_\mu\varphi^*+\epsilon^*_\alphadot F^*\,,\\
\delta F = -i \epsilon^*_\alphadot\sigmabar^{\mu\alphadot\alpha}\pd_\mu\eta\,, &~~~&\delta F^*=i\pd_\mu\eta^*\sigmabar^{\mu\alphadot\alpha}\epsilon_\alpha\,.\nonumber
\eeq
One can easily check that under (\ref{eq:SUSYtransforms}) the Lagrangian shifts by a total derivative, equations of motion are unaffected, and the 
transformation (\ref{eq:SUSYtransforms})
is a symmetry of the theory. 

A simple generalization of (\ref{eq:free}) leads to our first example of supersymmetry breaking. Since $F$ transforms into a total derivative, the Lagrangian remains supersymmetric after the addition of the term
\beq
\label{eq:freeF}
\Delta \Lagr = \mu^2 F+h.c. \,.
\eeq
On the other hand,
solving the $F$-term equation of motion we find $F=-\mu^2$. Substituting this result into (\ref{eq:SUSYtransforms}) we find that the ground state is not invariant and SUSY is spontaneously broken. 
We thus establish that $F$-term is an order parameter for SUSY breaking. This is also consistent with our earlier argument that the vacuum energy is such an order parameter because in our model $E=|F|^2$.

Finding SUSY breaking in a theory described by (\ref{eq:free}) and (\ref{eq:freeF})
may seem puzzling at first. Indeed, in a limit $\Mpl\rightarrow\infty$ we are still dealing with a non-interacting theory which only differs from our original example by a non-observable shift in zero-point energy. Inclusion of gravity, however, would resolve the puzzle by turning both examples into interacting theories with SUSY broken spontaneously in the second model. In a locally supersymmetric theory, $F$-terms (and $D$-terms which will be introduced later) still play the role of order parameters for SUSY breaking. On the other hand, the vacuum energy of a non-supersymmetric ground state is arbitrary while vacuum energy of a supersymmetric ground state is non-positive.

Having constructed our first (if trivial) example of a supersymmetric field theory, we would like to proceed to study interacting models. Given supersymmetry transformations (\ref{eq:SUSYtransforms}) this is relatively straightforward. 
Let's begin by adding an arbitrary interaction to the Lagrangian, for example, $-\varphi\eta\eta+\hc$ term. Its variation under SUSY transformations is given by:
\beq
\delta(-\varphi\eta\eta)=  2i \varphi \eta\pd_\mu\varphi \sigma^\mu\epsilon-2\varphi\eta\epsilon F\,,
\eeq
where we suppressed spinor indices.
To cancel this variation, additional Lagrangian terms are required. A reasonable guess is an interaction term $F\varphi^2+\hc$:
\beq
\delta(F\varphi^2)=-i \epsilon^*\sigmabar^{\mu}\pd_\mu\eta\varphi^2+2F\varphi\eta\epsilon\,.
\eeq
It is easy to see that the sum of the two variations is a total derivative and the theory is supersymmetric.
We can now write down the simplest interacting SUSY theory, an interacting Wess-Zumino model
\begin{equation}
\label{eq:WZcomp}
 \Lagr=|\pd_\mu\varphi|^2+i\eta^\dagger \pd_\mu\sigmabar^\mu \eta +|F|^2+(\lambda F\varphi^2-\lambda \varphi \eta\eta+\hc)\,,
\end{equation}
where $\lambda$ is a coupling constant.

\subsection{Superfield formalism}
It is possible to extend the procedure discussed in the previous section to more complicated theories but it becomes increasingly complicated, moreover some interactions can not appear in a supersymmetric Lagrangian.
Therefore it is useful to introduce a new formalism which will allow us to treat all superpartners as a single field (or {\em superfield}). Scalars and fermions related by supersymmetry should simply correspond to different components of a single superfield very much like spin up and spin down states are different components of a single fermion. To arrive at the desired superfield formalism it is convenient to introduce the notion of the superspace by extending 4 commuting spacetime coordinates $\{x_\mu\}$ to 4 commuting and 4 anti-commuting coordinates $\{x_\mu,\theta^\alpha, \thetabar_\alphadot\}$, where $\thetabar_\alphadot=(\theta^\alpha)^*$. The new coordinates satisfy anti-commutation relations
\beq
\{\theta_\alpha,\thetabar_\betadot\}=\{\theta_\alpha,\theta_\beta\}=\{\thetabar_\alphadot,\thetabar_\betadot\}=0\,.
\eeq

We can also define integrals over the superspace
\begin{equation}
\label{eq:sintegrals}
 \begin{aligned}
&\int d\theta=\int d\thetabar=\int d\theta\thetabar=\int d\thetabar \theta=0\,,\\
&\int d\theta^\alpha \theta_\beta=\delta^\alpha_\beta,~~~\int d\thetabar_\alphadot\thetabar^\betadot=\delta_\alphadot^\betadot\,,\\
& \int d^2\theta \theta^2=\int d^2\thetabar\thetabar^2 \,,\\
&\int d^4\theta\theta^2\thetabar^2=1\,,
 \end{aligned}
\end{equation}
where 
\begin{equation}
 \begin{aligned}
  &d^2\theta\equiv -\frac{1}{4}\epsilon_{\alpha\beta}d\theta^\alpha d\theta^\beta\,,\\
&d^2\thetabar\equiv -\frac{1}{4}\epsilon^{\alphadot\betadot}d\theta_\alphadot d\theta_\betadot\,,\\
&d^4\theta\equiv d^2\thetabar d^2\theta\,.
 \end{aligned}
\end{equation}

Functions of superspace coordinates are quite simple ---  the expansion in power series terminates at order $\theta^2\bar\theta^2$. 
Furthermore, integration and differentiation over superspace coordinates lead to the same results.

We can now express any supermultiplet as a single {\em superfield} which depends on superspace coordinates. Expanding in the Taylor series we have for the most general scalar superfield ({\em i.e.} superfield whose lowest component is a scalar field)
\beq
\label{eq:superfield}
\begin{aligned}
\Phi(x_\mu,\theta,\thetabar)=&\varphi(x_\mu) + \theta \eta(x_\mu)+\thetabar \chi^\dagger(x_\mu) 
+\thetabar \sigmabar^{\mu} \theta V_\mu(x_\mu)\\& + \theta^2 F(x_\mu)+\bar\theta^2 \bar F(x_\mu)+\ldots +\theta^2\thetabar^2D(x_\mu)\,.
\end{aligned}
\eeq
While $\Phi$ depends on a finite number of component fields, it has many more components than is necessary to reproduce the simplest free supersymmetric theory described in section \ref{sec:SUSY}. It turns out that the most general superfield (\ref{eq:superfield}) gives a {\em reducible} representation of supersymmetry. To describe the Wess-Zumino model (\ref{eq:WZcomp}), we will construct an irreducible representation of SUSY by imposing additional conditions on $\Phi$. To that end we will consider {\em chiral} and {\em antichiral} superfields $\Phi$ and $\Phi^\dagger$ respectively satisfying conditions
\beq
\bar D_{\dot\alpha} \Phi=0\,,~~~~D_{\alpha} \Phi^\dagger=0\,,
\eeq
where 
\beq
D_\alpha=\frac{\pd}{\pd\theta^\alpha}-i \sigma^\mu_{\alpha\alphadot}\thetabar^\alphadot \frac{\pd}{\pd_\mu}\,,~~~\bar D_\alphadot=-\frac{\pd}{\pd\thetabar^\alphadot}+i\theta^\alpha \sigma^\mu_{\alpha\alphadot}\frac{\pd}{\pd_\mu}\,.
\eeq
It is convenient to introduce new variables, $y^\mu=x^\mu+i \thetabar \sigmabar^\mu \theta$ and $y^{\mu\dagger}=x^\mu-i \thetabar \sigmabar^\mu \theta$.
Note that $\bar D_\alphadot y^\mu=D_\alpha y^{\dagger\mu}=0$. Therefore, a superfield defined by
\begin{equation}
\Phi(y^\mu) = \varphi(y^\mu)+\sqrt{2} \theta \eta(y^\mu)+\theta^2 F(y^\mu)
\end{equation}
is chiral, $\bar D_\alphadot \Phi(y^\mu)=0$, while its hermitian conjugate is antichiral.
Expanding $\Phi(y^\mu)$ in powers of superspace coordinates, we find
\begin{equation}
\begin{split}
\Phi=&\varphi(x)-i \theta \sigma^\mu \thetabar \pd_\mu\varphi(x)-\frac{1}{4}\theta^2\thetabar^2\pd^2\varphi(x)\\
+&\sqrt{2}\theta\eta+\frac{i}{\sqrt{2}}\theta^2\pd_\mu\eta\sigma^\mu\thetabar+\sqrt{2}\theta^2 F(x) \,.
\end{split}
\end{equation}
As we can see, chiral superfield only depends on three component fields and is a good candidate to describe our supersymmetric theory. 
To write supersymmetric Lagrangians using chiral superfields recall that  $F$ transforms into a total derivative and therefore all the $F$-terms in the Lagrangian are invariant under SUSY transformations. The supersymmetric Lagrangian term (\ref{eq:freeF}) can be written simply as
\beq
\Delta \mathcal{L}=\int d^2\theta \mu^2 \Phi+ \hc= \mu^2 F(x)+\mu^{\dagger 2} F^\dagger(x)\,.
\eeq
It is easy to check that any analytic  function of a chiral superfield, $W(\Phi)$, is also a chiral superfield and its $\theta^2$ component of  transforms into a total derivative. $W(\Phi)$, referred to as a superpotential, together with its hermitian conjugate $W(\Phi^\dagger)$ gives rise to supersymmetric interactions. After trivial generalization to a theory with several chiral superfields we can write:
\beq
\Lagr_W =\int d^2\theta\, W(\Phi_i)+\hc\,.
\eeq

Superpotential allows us to introduce a broad set of supersymmetric interactions. However, $W|_{\theta^2}$ does not contain spacetime derivatives, and therefore, does not lead to kinetic terms. It turns out that a $\theta^2\thetabar^2$ component of a real function of chiral superfields,  a \kahler potential, is also invariant under the supersymmetry transformations. In fact the simplest \kahler potential
gives rise to canonical kinetic terms:
\begin{equation}
\label{eq:kahler}
\begin{split}
 K=&\sum_i \Phi_i^\dagger \Phi_i\,,\\
\Lagr_K=&\int d^4\theta K= \sum_i\left(|\pd_\mu\varphi_i|^2+i\eta^*_i\pd_\mu\sigmabar^\mu\eta_i+ |F_i|^2\right)\,.
\end{split}
\end{equation}
A general \kahler potential leads to more complicated terms in the action \beq
\Lagr = \int d^4\theta K \supset g^{ij}(\pd_\mu \varphi^*_i\pd^\mu \varphi_j+i\eta_i^*\sigmabar^\mu\pd_\mu \eta_j + F^*_iF_j)\,,
\eeq
where $g^{ij}=\pd^2K/(\pd\Phi^\dagger_i\pd \Phi_j)|_{\Phi=\varphi}$ is a \kahler metric which implicitly depends both on the fields and parameters of the theory.
The \kahler metric determines the normalization of the kinetic terms and at a quantum level it  contains information about wave-function renormalization.

We are now ready to write down a general form of the Lagrangian in an interacting theory of chiral superfields
\begin{equation}
\label{eq:chiralL}
\begin{split}
\Lagr=&\int d^4\theta\, K(\Phi_i)+\int d^2\theta W(\Phi_i) + \int d^2\thetabar  W (\Phi_i^\dagger)\\
=&\, g^{ij}\left(\pd \varphi_i^* \pd \varphi_j + i \eta_i^*\pd_\mu\sigmabar^\mu \eta_j +F^*_iF_j\right)
 -\left(\frac{1}{2}\frac{\pd^2 W}{\pd \Phi_i \pd \Phi_j}\eta_i\eta_j-\frac{\pd W}{\pd \Phi_i}F_i+\hc\right)+\ldots\,,
\end{split}
\end{equation}
where dots represent possible higher order terms.
By solving $F$-term equation of motion we arrive at the scalar potential of the theory
\begin{equation}
\label{eq:scalarWZ}
 V=\frac{\pd \overline W}{\pd \Phi^\dagger_i}g_{ij}\frac{\pd W}{\pd \Phi_j}\,,
\end{equation}
where $g_{ij}=(g^{ij})^{-1}$.
We will generally assume that the \kahler potential is non-singular and therefore extrema of the  superpotential correspond to supersymmetric ground states of the theory. However, even in the supersymmetric vacuum information about the spectrum of the theory requires knowledge of the \kahler potential.

As an explicit example of the superfield formalism let us write down Lagrangian of interacting Wess-Zumino model \cite{Wess:1974jb}.
We will assume canonical \kahler potential (\ref{eq:kahler}) and the superpotential
\begin{equation}
 \label{eq:WZ-W}
W=\frac{m}{2}\Phi^2+\frac{\lambda}{3}\Phi^3\,.
\end{equation}
Lagrangian in terms of component fields takes the form
\begin{equation}
 \label{eq:WZ-m-component}
\Lagr=|\pd_\mu\varphi|^2+i\eta^\dagger \pd_\mu\sigmabar^\mu \eta +|F|^2+\left (mF\varphi +\lambda F\varphi^2-\frac{m}{2}\eta\eta-\lambda \varphi \eta\eta+\hc\right)\,.
\end{equation}

We now briefly discuss generalization to local supersymmetry or supergravity (SUGRA). In SUGRA the scalar potential becomes (neglecting $D$-terms that will appear in gauge theories):
\begin{equation}
\label{eq:VSUGRA}
V= \exp\left(\frac{K}{\Mpl}\right)\left((g^{ij}(D_iW)(D_jW)^*-\frac{3|W|^2}{\Mpl^2}\right)\,,
\end{equation}
where $D_i$ is a covariant supergravity derivative 
\begin{equation}
D_iW=\pd_i W+K_iW/\Mpl\,. 
\end{equation}
The $F$-type order parameters for SUSY breaking now involve covariant derivatives $F_i=D_i W$.
In supersymmetric vacua  $D_iW=0$ but as advertized earlier cosmological constant may be either  zero or negative depending on the vev of $W$. In phenomenological applications, one is interested in vacua with zero cosmological constant and broken supersymmetry --- this can always be achieved by shifting the superpotential by a constant, $W(\Phi_i)\rightarrow W(\Phi_i)+W_0$ and adjusting $W_0$ to cancel $D$- and $F$-term cotributions to vacuum energy.

Another important consequence of promoting supersymmetry to a local symmetry is the requirement that there exists a spin-3/2 superpartner of the graviton, gravitino. We will not write down the full gravitino  Lagrangian carefully but will note one important term
\begin{equation}
 \Lagr_\text{gravitino}\supset e \exp\left(\frac{K}{2\Mpl^2}\right) \left( \frac{\overline W}{\Mpl^2} \psi_\mu \sigma^{\mu\nu}\psi_\nu +\frac{W}{\Mpl^2} \psi^\dagger_\mu \sigmabar^{\mu\nu}\psi^\dagger_\nu \right)\,,
\end{equation}
where $e$ is a vierbein. We see that the gravitino is massive whenever the superpotential has non-vanishing vev. In particular, it is massive in supersymmetric vacua with negative cosmological constant --- but this is actually required by SUSY in anti-de Sitter spacetime. Supersymmetric vacua in Minkowski spacetime imply vanishing $\vev{W}$ and a massless gravitino. Once supersymmetry is broken and the vev of the superpotential is tuned to obtain the flat background, gravitino mass is determined by SUSY breaking parameters. In particular, when SUSY is broken by an $F$-term vev the gravitino mass becomes
\begin{equation}
 m_{3/2}^2=e^{K/\Mpl^2}\frac{F^*_i g^{ij}F_j}{3\Mpl^2}\,.
\end{equation}

\subsection{Vector Superfield}
Our next task is to construct a Lagrangian of a supersymmetric gauge theory. A gauge field $A_\mu$ has 2 on-shell degrees of freedom and 3 off-shell degrees of freedom. We already know that a Weyl fermion has 2 on-shell degrees of freedom, and thus is a good candidate for being the superpartner of the gauge boson, gaugino. Off-shell, however, degrees of freedom do not match. Just as in the case of a chiral superfield, we need to introduce an auxiliary scalar field with one off-shell degree of freedom. This field, denoted by $D$, must be a real scalar without a kinetic term.

SUSY transformations for the components of the vector multiplet are given by
\begin{equation}
\label{eq:SUSYtransformV}
\begin{split}
  \delta A^a_\mu=&-\frac{1}{\sqrt{2}}\left(\epsilon^\dagger \sigmabar_\mu \lambda^a+\lambda^{\dagger a}\sigmabar^\mu\epsilon\right)\,,\\
  \delta \lambda^a_\alpha=&-\frac{1}{2\sqrt{2}}\left(\sigma^\mu\sigmabar^\nu\epsilon \right)_\alpha F^a_{\mu\nu}+\frac{1}{\sqrt{2}}\epsilon_\alpha D^a\,,\\
  \delta \lambda^{\dagger a}_\alpha=&-\frac{1}{2\sqrt{2}}\left(\epsilon^\dagger \sigmabar^\mu\sigma^\nu\right)_\alphadot F_{\mu\nu}^a+\frac{1}{\sqrt{2}} F^a_{\mu\nu}+\frac{1}{\sqrt{2}}\epsilon^\dagger_\alphadot D^a\,,\\
 \delta D^a=&-\frac{i}{\sqrt{2}}\left(\epsilon^\dagger \sigmabar^\mu D_\mu \lambda^a-D_\mu\lambda^{\dagger a}\sigmabar^\mu \epsilon\right)\,.
\end{split}
\end{equation}

To give a superfield description of a vector multiplet containing $A_\mu$, $\lambda$, and $D$ consider a real superfield
\beq
\label{eq:vectorcond}
V=V^\dagger\,.
\eeq
In components this superfield can be written as
\begin{equation}
\begin{split}
V=&\frac{1}{2} C+i\theta\chi +\frac{i}{2}\theta^2(M+iN)
+\theta\sigma^\mu\thetabar A_\mu \\&+i\theta^2\thetabar(\lambda^\dagger-\frac{i}{2}\sigmabar^\mu\pd_\mu\chi)
+\frac{1}{4}\theta^2\thetabar^2(D(x)+\frac{1}{2}\Box C(x))+\hc\,.
\end{split}
\end{equation}
The vector superfield contains more degrees of freedom than we hoped for. However, these additional degrees of freedom are auxiliary fields required by gauge invariance in a theory of the massless vector field that we wish to formulate.
To see this, introduce a chiral superfield $\Lambda$:
\begin{equation}
 \Lambda= \frac{\alpha(y_\mu)+i\beta(y_\mu)}{2}+ \theta\frac{\xi(y_\mu)}{\sqrt{2}}+\frac{1}{2}\theta^2 f(y_\mu)\,.
\end{equation}
If we shift the vector superfield according to
\begin{equation}
\label{eq:Uonesusygauge}
V\rightarrow V+i(\Lambda-\Lambda^\dagger)\,.
\end{equation}
we find that its vector component is gauge transformed:
\begin{equation}
 A_\mu\rightarrow A_\mu+\pd_\mu \alpha
\end{equation}
and the shift by $\Lambda$ represents a SUSY generalization of a regular gauge transformation.
Other components of a vector superfield transform according to 
\begin{equation}
\begin{split}
 C&\rightarrow C -\beta\,,\\
  \chi&\rightarrow \chi+ \xi\,,\\
 M+iN&\rightarrow M+iN+2f\,,\\
 \lambda&\rightarrow\lambda\,,\\
 D&\rightarrow D\,.\\
\end{split}
\end{equation}
We would like to maintain gauge invariance as an explicit symmetry of the Lagrangian, however we can use the remaining components of $\Lambda$ to set all the auxiliary fields other than $D$ in a vector multiplet to zero. This is equivalent to a gauge choice and is referred to as Wess-Zumino gauge. The Wess-Zumino gauge is very convenient in practice despite the fact that in this gauge SUSY is not fully manifest.

\subsection{Supersymmetric $U(1)$ gauge theory}
\label{sec:uone}
To write down kinetic terms for the vector superfields we define a chiral spinor superfield
\beq
\Wg_\alpha=-\frac{1}{4}\bar D^2 D_\alpha V\,.
\eeq
As usual, analytic functions of chiral superfields are themselves chiral superfields.
Therefore, $\theta^2$ component of $\Wg^\alpha \Wg_\alpha$ transforms into a total derivative. 
In additon $\Wg^\alpha \Wg_\alpha$ contains gauge kinetic terms allowing us to write supersymmetric Lagrangian for  $U(1)$ theory as
\begin{equation}
\begin{split}
 \mathcal{L}=& \left(\int d^2\theta\frac{1}{4g^2} \Wg^\alpha \Wg_\alpha+\int d^2\thetabar \frac{1}{4g^2} \Wg^\dagger_\alphadot \Wg^{\dagger\alphadot}\right)\\
=&\frac{1}{4g^2} F_{\mu\nu}F^{\mu\nu}+\frac{i}{g^2}\lambda^\dagger\pd_\mu\sigmabar^\mu\lambda+\frac{1}{2g^2}D^2\,.
\end{split}
\end{equation}
Notice that we have chosen a non-canonical normalization for kinetic terms --- as we will see later, this normalization is very convenient in SUSY gauge theories.

A chiral superfield of charge $q$ transforms under gauge transformations according to
\begin{equation}
\Phi \rightarrow e^{-iq\Lambda}\Phi\,.
\end{equation}
A \kahler potential of the form
\begin{equation}
 K= K\left(\Phi^\dagger, e^{qV} \Phi\right)\,\\
\end{equation}
is gauge invariant.
In particular, canonical \kahler potential 
\begin{equation}
  K= \Phi^\dagger e^{qV} \Phi\,
\end{equation}
contains regular gauge interactions as well as new interactions of matter fields and gauginos
\begin{equation}
 -\sqrt{2}\left(\phi^* \eta\lambda+\lambda^\dagger\eta^\dagger \phi\right)\,.
\end{equation}

Combining this with kinetic terms for vector superfield and introducing a general gauge invariant superpotential we can write the Lagrangian for a theory with several charged chiral multiplets
\begin{equation}
 \Lagr=\left( \int d^2\theta \frac{1}{4g^2} \Wg^\alpha \Wg_\alpha + \hc\right) + \int d^4\theta \sum_i \Phi_i^\dagger e^{q_iV} \Phi_i^\dagger +\left(\int d^2\theta W(\Phi_i)+\hc\right) \,.
\label{eq:u1sym}
\end{equation}
Integrating over the superspace coordinates leads to the component Lagrangian. It is especially useful to look at D-terms:
\begin{equation}
 \Lagr_D=\frac{1}{2g^2}D^2+D\left(\sum_i q_i|\varphi_i|^{2}\right)^2\,.
\end{equation}
Integrating out D-term we find a new contribution to the scalar potential
\begin{equation}
 V_D=\frac{g^2}{2}\left(\sum_i q_i|\varphi_i|^2\right)^2\,.
\end{equation}
It is easy to see that anomaly free $U(1)$ gauge theories necessarily have directions in the field space along which $V_D$ vanishes. These directions are referred to as $D$-flat. As we will see shortly, some chiral non-abelian gauge theories do not have $D$-flat directions. Moreover, some of the D-flat directions may be lifted by the F-term potential $V_F$. However, it is quite generic in SUSY gauge theories that there exist directions in the field space along which both $V_D$ and $V_F$ vanish. Submanifold of the field space satisfying both $D$- and $F$-flatness conditions represents (classical) vacua in the theory and is often referred to as a (classical) {\em moduli space}. Field fluctuations along the moduli space are massless and are called moduli. In section \ref{sec:SUSYQCD} we will see that classical moduli space may be modified or completely lifted by non-perturbative dynamics, however, even then it plays a useful role in the analysis of dynamical properties of the theory.

Before moving on to a discussion of non-abelian gauge theories, we should consider one more supersymmetric and gauge invariant term, the Fayet-Illiopoulus D-term, that can arise only in an abelian case. According to (\ref{eq:Uonesusygauge}) vector superfield $V$ is not invariant under SUSY transformations nor are its components  invariant under $U(1)$ gauge transformation. However, both $D$ and $\lambda$ are neutral under the gauge symmetry. Furthermore, the $\theta^4$ component of the vector superfield is a scalar and shifts only by a total derivative. Therefore the following SUSY and gauge invariant term can be added to the Lagrangian
\begin{equation}
\int d^4\theta \xi^2 V\,.
\label{eq:Dterm}
\end{equation}

Upon integrating out an auxiliary D-term, the D-term potential in the theory becomes
\begin{equation}
 V_D=\frac{g^2}{2}\left(\sum_i q_i|\varphi_i|^2+\xi^2\right)^2\,.
\label{eq:DtermV}
\end{equation}

\subsection{Non-abelian gauge theory}
\label{sec:nonabelian}
It is easy to extend our discussion to the case of non-abelian gauge theories. Gauge transformation take the form
\begin{equation}
 V\rightarrow e^{-i \Lambda}Ve^{i \Lambda}\,,
\end{equation}
where $V$ transforms in an adjoint representation of the gauge group and $\Lambda=\Lambda^aT^a$.

The Lagrangian for a supersymmetric Yang-Mills theory with matter fields is a simple generalization of (\ref{eq:u1sym}).
In particular, the gauge kinetic terms become
\begin{equation}
 \Lagr_\text{SYM}=\int d^2\theta \frac{1}{2g^2} \Tr \Wg^\alpha \Wg_\alpha + \hc\,.
\end{equation}

To discuss physics of the theory with matter fields, it is convenient to consider a specific example of an  $SU(N)$ gauge theory with  $F$ pairs of chiral superfields in fundamental and anti-fundamental representations, $Q$ and $\bar Q$ respectively. We will refer to this matter content as $F$ flavors. Quantum numbers of the matter fields under local and non-anomalous global symmetries are given by
\begin{equation}
\label{eq:sunf}
\renewcommand{\arraystretch}{1.5}
 \begin{array}{|c|c|cccc|}
\hline
&SU(N)&SU(F)_L&SU(F)_R&U(1)_B&U(1)_R\\
\hline
Q&N&F&1&1&\frac{N_f-N_c}{N_f}\\
\bar Q&\bar N&1&\bar F&-1&\frac{N_f-N_c}{N_f}\\
\hline
   \end{array}
\end{equation}

The D-term potential has the form
\begin{equation}
 V_D=\frac{g^2}{2}\left( q^\dagger T^a q - \bar q T^a \bar q^\dagger\right)^2\,
\end{equation}
and has many classical flat directions. It is possible to show  that it vanishes when squark vevs satisfy the following condition \cite{Affleck:1983mk,Affleck:1984xz}
\begin{equation}
 q^{\dagger if} q_{jf}-\bar q^{if}\bar q^\dagger_{jf}=\alpha \delta^i_j\,,
\end{equation}
where $i$ and $f$ are color and flavor indices respectively and $\alpha$ is an arbitrary constant.
In a theory with $F<N$ one can use gauge and global symmetry transformations to write the vevs in the form 
\begin{equation}
\label{eq:flatfln}
 Q=\bar Q=\left(\begin{array}{cccccc}
                 v_1&&&0\\
                 &v_2&&\\
                 &&\ddots&\\
                 &&&v_F\\
                 &   \\
                 0&\cdots&&0
                \end{array}\right)\,.
\end{equation}
When $F\ge N$ flat directions can be parameterized by
\begin{equation}
\label{eq:flatfmn}
 Q=\left(\begin{array}{cccccc}
                 v_1&&&&&0\\
                 &v_2&&&&\vdots\\
                 &&\ddots&&&\\
                 0&\cdots&&v_F&&0
                \end{array}\right)\,,
~~~~
Q=\left(\begin{array}{cccccc}
                 \bar v_1&&&&&0\\
                 &\bar v_2&&&&\vdots\\
                 &&\ddots&&&\\
                 0&\cdots&&\bar v_F&&0
                \end{array}\right)\,,
\end{equation}
where $|v_i|^2-|\bar v_i|^2=\alpha$.

An equivalent description of the moduli space can be obtained if we note that all states related to (\ref{eq:flatfln}) by global symmetry transformations may be parameterized in terms of gauge invariant composites 
\begin{equation}
\label{eq:meson}
M_{ff^\prime}=Q_f\bar Q_{f^\prime}\,. 
\end{equation}
 We will refer to $M_{ff^\prime}$ as mesons. Additional flat directions arising when $F\ge N$ can be described in terms of baryons and anti-baryons
\begin{equation}
\label{eq:baryon}
B_{f_{N+1}\ldots f_F}=\epsilon_{f_1\ldots f_F}\epsilon^{a_1\ldots a_n}Q_{a_1}^{f_1}\ldots Q_{a_N}^{f_N},~~~
\bar B_{f_{N+1}\ldots f_F}=\epsilon_{f_1\ldots f_F}\epsilon^{a_1\ldots a_n}\bar Q_{a_1}^{f_1}\ldots \bar Q_{a_N}^{f_N}\,.
 \end{equation}
Not all of these gauge invariant composites are independent. When $F=N$ we have a classical relation
\begin{equation}
\label{eq:feqnconstraint}
 \det M=B\bar B\,.
\end{equation}
Similarly, for $F>N$
\begin{equation}
\begin{split}
 \det M\, M^{-1}_{ff^\prime}=B_f\bar B_f^\prime\,,\\
 B^f M_{ff^\prime}=M_{ff^\prime} \bar B^{f^\prime}=0\,.
\end{split}
\end{equation}

\section{Non-renormalization theorems}
\label{sec:NRtheorem}

Clearly the presence of an additional symmetry must simplify calculation of quantum corrections to the Lagrangian. At the very least SUSY requires that counterterms for interactions related by symmetry are identical. For example, in the Wess-Zumino model (\ref{eq:WZ-m-component}) counterterms for both  $F\varphi^2$ and $\varphi\eta\eta$ must be the same. Similarly, in SUSY gauge theories renormalization of the gauge coupling, gaugino-scalar-fermion coupling, and quartic scalar coupling in the $D$-term potential must be related.
However, it turns out that SUSY imposes much more powerful constraints on supersymmetric Lagrangians. 
Namely, only \kahler potential terms are renormalized to all orders in perturbation theory. On the other hand, superpotential terms are not renormalized while gauge coupling and the Fayet-Illiopolous $D$-term are not renormalized beyond one loop order. One can prove these statements, known as non-renormalization theorems, order by order in perturbation theory \cite{Grisaru:1979wc}. Instead we will consider a much slicker derivation due to Seiberg \cite{Seiberg:1993vc} which uses the symmetries and holomorphy of SUSY Lagrangians. This derivation will also allow us to see how  non-perturbative dynamics may  affect low energy physics of SUSY gauge theories.

\subsection{R-symmetry}
An important role in our discussion will be played by an $R$-symmetry --- a symmetry that rotates\\\\\\\\\
 superspace coordinates $\theta$ by a phase, $\theta\rightarrow e^{i\alpha} \theta$. Defining the R-charge of $\theta$ to be $R_\theta=1$ and using the equation (\ref{eq:sintegrals}) we find $R_{d\theta}=-1$. Lagrangian terms arising from the \kahler potential are always invariant under the $R$-symmetry since both $d^4\theta$ and the \kahler potential are real. On the other hand, the invariance of the full Lagrangian is neither guaranteed nor required. $R$-symmetry is only a symmetry of the Lagrangian if the superpotential transforms with the charge 2, $W\rightarrow e^{2i\alpha} W$. Imposing these transformation properties on the superpotential determines R-symmetry charges of the superfields. It is possible that a consistent assignment of R-charges does not exist and then R-symmetry is explicitly broken by some terms in the Lagrangian.

For example, consider an interacting massless Wess-Zumino model. If we assign an $R$-charge of $2/3$ to $\Phi$, the superpotential has $R$-charge 2 and the Lagrangian is invariant under $R$-symmetry. Alternatively, we can consider free theory with a non-vanishing mass. In this case an $R$-charge of $\Phi$ under $R$-symmetry must be  1. On the other hand, in a massive interactive Wess-Zumino model there is no charge assignment which leaves the Lagrangian invariant under R-symmetry.

R-symmetry is quite unusual. Unlike all other global symmetries, it acts on superspace coordinates $\theta$ and $\thetabar$. As a result different components of the superfields transform differently under R-symmetry. Consider, for example, a chiral superfield with R-charge $R$. Its lowest component has the same R-charge as the superfield itself, its $\theta$ component has R-charge $R-1$ and its $\theta^2$ component has $R$-charge $R-2$.

R-charges of matter fields may depend on the model under consideration. However, R-charges of the fields in a vector multiplet are uniquely fixed. Indeed $\Wg^\alpha\Wg_\alpha$ has R-charge 2. Since gaugino $\lambda$ is the lowest component of $\Wg^\alpha$ its R-charge is $1$ while the $D$-term and the gauge field $A_\mu$ are neutral (as should have been expected for real fields).

\subsection{Superpotential terms}
To prove non-renormalization theorems we will use all the symmetries of the SUSY field theories, including an $R$-symmetry. Moreover, we will be able to use $R$-symmetry even in models where it is explicitly broken by the superpotential interactions. To this end we will  promote the parameters of the Lagrangian to background superfields \cite{Seiberg:1993vc}. Consider, for example, the Wess-Zumino model (\ref{eq:WZ-W}).
We will interpret this model as an effective low energy description of a more fundamental theory in which parameters $m$ and $\lambda$ arise as vacuum expectation values of heavy superfields. This interpretation enhances apparent symmetries of the theory. The model now has a $U(1)\times U(1)_R$ global symmetry which is spontaneously broken by expectation values of the spurions $m$ and $\lambda$.
The charges of the dynamical superfield $\Phi$ and spurions under the symmetries of the theory are given by
\beq
\begin{array}{|c|c|c|}
\hline
 &U(1)_R&U(1)\\
\hline
\Phi&1&1\\
m&0&-2\\
\lambda&-1&-3\\\hline
\end{array}
\eeq

In this approach superpotential must be described by a holomorphic function of both the dynamical and background superfields. On the other hand, the \kahler potential is still a general real function of superfields and spurions:
\begin{equation}
\begin{split}
K&=K(\Phi^\dagger, \Phi, m^\dagger, m, \lambda^\dagger, \lambda)\,,\\
W&=W(\Phi, m,\lambda)\,.
\end{split}
\end{equation}

The requirement that the renormalized superpotential is holomorphic and has correct transformation properties under global symmetries  restricts its form to be
\beq
W=\frac{m}{2}\Phi^2 f\left(\frac{\lambda\Phi}{m}\right)\,.
\eeq

In the weak coupling limit the effective superpotential should approach the classical one and therefore there should exist a Taylor series expansion of $f$ in  $\lambda\Phi/m$: \begin{equation}
W=\frac{m}{2}\Phi^2\left(1+\frac{2}{3}\frac{\lambda\Phi}{m}+\Order{\frac{\lambda^2\Phi^2}{m^2}}\right)=
\frac{m}{2}\Phi^2+\frac{\lambda}{3}\Phi^3 +\Order{\lambda^2}\,.
\end{equation}
Thus
\beq
f\left(\frac{\lambda\Phi}{m}\right)=1+\frac{2}{3}\frac{\lambda\Phi}{m}+\Order{\lambda^2}\,.
\eeq
Furthermore, the $m\rightarrow 0$ limit must be regular, and therefore $W$ should not contain negative powers of $m$. Thus (\ref{eq:WZ-W}) is exact \cite{Seiberg:1993vc}. No higher dimension terms are generated. This, in particular, means that there are no counterterms leading to renormalization of $m$ or $\lambda$. To make the latter conclusion obvious, let's assume, for example, that a counterterm $\delta_m$ is generated at some order in the perturbation theory. Then $\delta_m \sim \lambda^n$ --- but as we just argued no such powers of $\lambda$ can appear in the superpotential.

The \kahler potential $K$, on the other hand, is real and can be a function of $|m|^2$ as well as of $|\lambda|^2$ both of which are invariant under all the symmetries. After renormalization \kahler potential becomes
\begin{equation}
K=\Phi^\dagger\Phi \rightarrow Z\left(\frac{\mu^\dagger\mu}{m^\dagger m},\lambda^\dagger\lambda\right)\Phi^\dagger\Phi\,. 
\end{equation}
The renormalization of the \kahler potential implies that the physical coupling constants,
 $m$ and $\lambda$, are renormalized in contrast to holomorphic parameters in the superpotential. Nevertheless an RG evolution of coupling constants is completely determined by wave function renormalization.

\subsection{Gauge coupling renormalization}
To discuss the renormalization of the gauge coupling constant in SUSY gauge theories we will once again  promote the gauge coupling function to a background superfield whose lowest component is 
\beq
\tau=\frac{8\pi^2}{g^2}+i\theta_\text{YM}\,.
\eeq
The gauge field Lagrangian then becomes
\begin{equation}
\frac{1}{4g^2}\int d^2\theta \Wg^\alpha \Wg_\alpha +\hc \rightarrow \frac{1}{32\pi^2}\int d^2\theta\, \tau\Wg^\alpha \Wg_\alpha +\hc\supset \frac{1}{4g^2}F^2_{\mu\nu}+\frac{\theta_{YM}}{32\pi^2}F\tilde F\,.
\end{equation}
$F\tilde F$ is a total derivative and does not affect local equations of motion. An abelian theory is invariant under shifts $\theta_\text{YM}\rightarrow \theta_\text{YM}+const$.  In a non-abelian theory the action must be  a periodic function of $\theta_\text{YM}$ with a period $2\pi$.
To see that recall that in a non-abelian theory there exist topologically non-trivial gauge configurations whose contribution to the action is
\begin{equation}
\frac{\theta_{YM}}{32\pi^2}\int d^4 x F\tilde F = n \theta_{YM}\,,
\end{equation}
where $n$ is a winding number of the field configuration. To calculate correlation functions one needs to sum over all $n$ and periodicity in $\theta_{\mathrm{YM}}$ follows immediately. 

These arguments imply that renormalization can at most shift the coefficient of the gauge kinetic term by a constant which corresponds to one-loop renormalization. No higher order corrections are allowed. 
The one loop coefficient of the $\beta$-function is given by
\beq
b_0=3C(G)-\sum_rC_2(r)\,.
\eeq
As an example, consider an $SU(N)$ SUSY gauge theory with $F$ flavors. In this theory renormalization group  evolution of a {\em holomorphic} gauge coupling is given by
\begin{equation}
\label{eq:holomorphicbeta}
\begin{split}
&b_0=3N-F\,,\\
&\frac{8\pi^2}{g^2(\mu)}=\frac{8\pi^2}{g^2(M)}+b_0 \ln\frac{\mu}{M}\,.
\end{split}
\end{equation}
However, we have seen in the Wess-Zumino model that even as holomorphic parameters are not renormalized, physical coupling constants are renormalized to all orders in the perturbation theory due to the wave function renormalization.
Similarly, physical gauge couplings in supersymmetric gauge theories are renormalized to all orders in  the perturbation theory. However, any renormalization beyond one loop is due to wavefunction renormalization and  
an exact $\beta$-function can be written in terms of anomalous dimensions of the matter fields \cite{Novikov:1983uc, Shifman:1986zi}:
\begin{equation}
\beta(\alpha)\equiv\frac{d\alpha(\mu)}{d\ln\mu}=-\frac{\alpha^2}{2\pi}\frac{3C(G)-
\sum_r C_2(r)(1-\gamma_r)}{1-\frac{\alpha}{2\pi}C(G)}\,,
\end{equation}
where
\begin{equation}
  \gamma_r=\frac{\pd \ln Z_r(\mu)}{\pd \ln \mu}\,.
\end{equation}

\subsection{D-term renormalization}
We conclude the discussion of non-renormalization theorems by considering Fayet-Illiopoulos $D$-terms (\ref{eq:Dterm}).
It was shown in \cite{Fischler:1981zk} that D-term is renormalized at most at one loop.
Once again spurion formalism is the most straightforward way to derive this result \cite{Dine:1996ui} (see also \cite{Shifman:1986zi}). 
Recall that while $D$-term of the $U(1)$ vector superfield is invariant both under gauge and supersymmetry transformations, the full superfield $V$ is not. 
If the D-term coefficient $\xi$ depends on coupling constants in the theory, it becomes superspace valued once we promote couplings to spurions. Then performing superspace integral in (\ref{eq:Dterm}) results in gauge non-invariant terms in the Lagrangian.
Therefore, $\xi$ must be a pure number independent of all the coupling constants in the theory. 
There are, however, one loop quadratically divergent diagrams generating a tadpole for $D$. These arise from the D-term coupling to matter superfields in (\ref{eq:DtermV}) and individually are quadratically divergent. The result, however, is only non-vanishing if the sum of $U(1)$ charges in a theory is non-vanishing, {\em i.e.} in theories where low energy Lagrangian suffers from gravitational anomalies.

\section{Non-perturbative dynamics in SUSY QCD}
\label{sec:SUSYQCD}
\subsection{Affleck-Dine-Seiberg superpotential}
Our discussion of non-renormalization theorems  made use of a requirement that the description of the theory is non-singular in weak coupling and massless limits. This immediately lead to a constraint that the superpotential does not contain negative powers of light superfields. 
While Lagrangian terms with negative powers of fields are never generated in the perturbation theory, it is known that such terms may arise due to  non-perturbative dynamics. Therefore, non-renormalization theorems may be violated by non-perturbative effects.

As an example consider $SU(N)$ SUSY gauge theory with $F$ flavors of matter fields in a fundamental representation introduced in section \ref{sec:nonabelian}. When $F<N$ symmetries of (\ref{eq:sunf}) allow the appearance of non-perturbative superpotential \cite{Davis:1983mz,Affleck:1984xz,Affleck:1983mk}
\begin{equation}
\label{eq:ADSW}
 W=C_{N,F}\left(\frac{\Lambda^{3N-F}}{\det{Q\bar Q}}\right)^{\frac{1}{N-F}}\,.
\end{equation}
The theory is strongly coupled near the origin of the moduli space and in general the coefficient $C_{N,F}$ is not calculable. However, if $C_{N,F}$ is non-zero, the superpotential (\ref{eq:ADSW}) forces $Q$ and $\bar Q$ to run away and $SU(N)$ is broken to an $SU(N-F)$ subgroup. There are no light charged matter fields left in the low energy physics (all the components of $Q$ and $\bar Q$ charged under unbroken group are eaten by the super-Higgs mechanism). The low energy effective field theory is described by a pure super Yang-Mills theory. It is expected that non-perturbative dynamics in pure SYM leads to a gaugino condensate $\vev{\lambda\lambda}=\Lambda_L^3$. 
To verify that (\ref{eq:ADSW}) is consistent with this expectation let us consider the evolution of holomorphic gauge coupling constants in the low and high energy theory and require that they match at the scale of gauge boson masses. Denoting this scale by $v\sim \sqrt{Q\bar Q}$ we can write
\begin{equation}
 \frac{8\pi^2}{g_L^2(\mu)}=\frac{1}{g_H^2(\Lambda_\text{UV})}+(3N-F)\ln\left(\frac{v^2}{\Lambda_\text{UV}}\right)-3(N-F)\ln\left(\frac{\mu^2}{v^2}\right)\,.
\end{equation}
This allows us to match the dynamical scales of two theories
\begin{equation}
\label{eq:lambdamatch}
 \Lambda_L^{3(N-F)}=\mu^{3(N-F)}\exp\left(-\frac{8\pi^2}{g^2_L(\mu)}\right)=\frac{\Lambda_\text{UV}^{3N-F}\exp\left(-\frac{8\pi^2}{g_H^2(\Lambda_\text{UV})}\right)}{v^{2F}}=\frac{\Lambda_H^{3N-F}}{v^{2F}}\,.
\end{equation}
We see that superpotential (\ref{eq:ADSW}) can be expressed in terms of the parameters of low energy physics and has the form expected of gaugino condensate
\begin{equation}
\label{eq:gauginocond}
 W=C_{N-F,0}\Lambda_L^3\,.
\end{equation}
Moreover we conclude that $C_{N,F}=C_{N-F,0}=C_{N-F}$. 

This argument provides non-trivial evidence that the superpotential (\ref{eq:ADSW}) is non-vanishing, however, we still have not calculated $C_{N-F}$.
Fortunately, there is one case where $C_{N-F}$ is calculable. In a model with $F=N-1$ the gauge group completely broken at a generic point on the classical moduli space. This allows one to perform an explicit  instanton calculation \cite{Affleck:1984xz,Affleck:1983mk,Cordes:1985um} and find
$C_{N,N-1}=C_1=1$. 

We can now derive $C_{N-F}$  for other values of $F$.
To do so, let us add mass term for one quark flavor
\begin{equation}
 W=C_{N-F}\left(\frac{\Lambda^{3N-F}}{\det Q\bar Q}\right)^{\frac{1}{N-F}}+m Q_F\bar Q_F\,.
\end{equation}
When $m\gg\Lambda$, heavy superfields decouple and low energy physics is described by a theory with $F-1$ flavors. Solving $Q_F$ and $\bar Q_F$ equations of motion we obtain an effective superpotential
\begin{equation}
\label{eq:ADSWFlN}
 W=C_{N-(F-1)}\left(\frac{m\Lambda^{3N-F}}{\det^\prime Q\bar Q}\right)^{\frac{1}{N-(F-1)}}\,,
\end{equation}
where prime implies that the determinant is taken only over $F-1$ light flavors and $C_{N-(F-1)}$ is determined is determined by $C_{N-F}$, $N$, and $F$.

Similarly to (\ref{eq:lambdamatch}) we can match the dynamical scales of two effective descriptions
\begin{equation}
 \Lambda^{3N-(F-1)}_L=m\,\Lambda^{3N-F}\,.
\end{equation}
We thus establish that (\ref{eq:ADSWFlN}) is indeed equivalent to (\ref{eq:ADSW}). Finally, the knowledge of $C_1$ allows us to find $C_{N-F}=N-F$.
 
To conclude our discussion of theories with $F<N$ we note that when all flavors are massive, the supersymmetric ground state is found at finite vevs and is given by
\begin{equation}
 \vev{Q_i\bar Q_j}=\left(\det m\, \Lambda^{3N-F}\right)^{1/N} m^{-1}_{ij}\,.
\end{equation}

\subsection{Quantum modified moduli space}
When the number of colors equals the number of flavors, there is no obvious superpotential that can arise dynamically. However, one can still perform an instanton calculation and find that the classical constraint (\ref{eq:feqnconstraint}) is modified \cite{Seiberg:1994bz}:
\begin{equation}
\label{eq:feqnquantum}
 \det M-B\bar B=\Lambda^{2N}\,.
\end{equation}
To interpret this result let us recall that in the classical theory the \kahler potential becomes singular at the origin of the field space, indicating the appearance of additional massless degrees of freedom --- gauge bosons of the restored $SU(N)$ symmetry. Non-perturbative effects lead to a spectacular result  --- the origin of the field space does not belong to the moduli space and the \kahler potential is non-singular everywhere. The gauge symmetry is not restored anywhere on the moduli space and the fluctuations of $M$, $B$, and $\bar B$ satisfying the constraint (\ref{eq:feqnquantum}) are the only massless particles in the quantum theory. 

It is convenient to parameterize the dynamics that led to (\ref{eq:feqnquantum}) by introducing an auxiliary Lagrange multiplier superfield $A$ and the superpotential
\begin{equation}
\label{eq:feqnW}
 W=A(\det M -B\bar B-\Lambda^{2N})\,.
\end{equation}
There are several checks that can be performed to verify that this superpotential leads to the correct description of physics. For example, adding a mass term for a single flavor results in a low-energy theory with $N-1$ flavors. Using (\ref{eq:feqnW}) and integrating out heavy flavor we indeed obtain the ADS superpotential (\ref{eq:ADSW}) appropriate for this theory.

\subsection{s-confinement}
\label{sec:sconfine}
Adding one more flavor to the theory leads to a model which exhibits confinement without chiral symmetry breaking \cite{Seiberg:1994bz}. In this model the classical constraint is not modified quantum mechanically. The low energy effective field theory can be described by the superpotential
\begin{equation}
\label{eq:sconfine}
 W=\frac{1}{\Lambda^{2N+1}}\left(BM\bar B - \det M\right)\,.
\end{equation}
All the degrees of freedom in $M$, $B$, and $\bar B$ are physical. At the origin of the field space the full global symmetry is restored and all components of $M$, $B$, and $\bar B$ become massless. This is in contrast to the interpretation of the singularity  in the classical description where the origin of the field space corresponds to an appearance of massless gluons.

 The validity of this description is supported by the fact that 't Hooft anomaly conditions match between the UV and IR descriptions for the full global symmetry. One can also verify the validity of the description by perturbing the theory with a mass term --- such a perturbation leads to a theory with the  quantum modified moduli space we described earlier.

\subsection{Dualities in SUSY QCD}
The story becomes even more interesting as we further increase the number of flavors, $F>N+1$. In this regime the infrared physics of an (electric) gauge theory has a dual description in terms of a magnetic theory with the same global symmetries but different gauge symmetry. It is convenient to start our discussion with an example of large $N$ and $F$ with $F<3N$. We will choose $\epsilon=1-F/(3N)$ to be positive and small.
In this case, the one loop $\beta$-function coefficient is small and the interplay between one and two loop running leads to a weakly coupled infrared fixed point \cite{Banks:1981nn}.
Seiberg \cite{Seiberg:1994pq,Seiberg:1994bz} argued that the physics at the ifrared fixed point has a dual (magnetic) description in terms of $SU(\tilde N)$ gauge group with $\tilde N=F-N$ colors. The dual theory has the same global symmetries and contains dual quarks, $q$ and $\bar q$, as well as elementary mesons $\widetilde M$ related to the gauge invariant composites of the electric description by
\begin{equation}
\widetilde M\sim \frac{1}{\mu}M=\frac{1}{\mu}(Q\bar Q)\,,
\end{equation}
where $\mu$ is the matching scale of the two descriptions.
In addition, there exist a correspondence between the baryon operators in two theories
\begin{equation}
\begin{aligned}
 B^{i_i\ldots i_N}\sim \epsilon^{i_1\ldots i_N j_1\ldots j_{\tilde N}}b_{j_1\ldots j_{\tilde N}}\,,\\
 \bar B^{i_i\ldots i_N}\sim \epsilon^{i_1\ldots i_N j_1\ldots j_{\tilde N}}\bar b_{j_1\ldots j_{\tilde N}}\,.\\
\end{aligned}
\end{equation}
If the superpotential term 
\begin{equation}
 W=\tilde M q\bar q \sim\frac{1}{\mu} M q\bar q
\end{equation}
is added to the Lagrangian, the magnetic theory flows to the (strongly interacting) infrared fixed point that is identical to the fixed point of the electric theory. 
The dynamical scales of two descriptions are related by
\begin{equation}
 \Lambda^{3N-F}\tilde \Lambda^{3\tilde N-F}=(-1)^{F-N} \mu^F\,.
\end{equation}

It is instructive to consider various deformations of the two theories. 
We will perturb the electic theory by the mass term for one flavor of quarks. The number of quark flavors in low energy physics is reduced by one and the theory flows to a new (slightly more strongly coupled) infrared fixed point.
In the magnetic description the perturbation corresponds to adding a tadpole for the meson field
\begin{equation}
 W=m Q\bar Q=m M \sim m\mu \widetilde M\,.
\end{equation}
This forces one flavor of dual quarks to acquire a vev, breaking the magnetic gauge group to $SU(\tilde N-1)$ and reducing the number of flavors by one. The magnetic theory flows to a new (less strongly coupled) infrared fixed point. Infrared physics of the two descriptions remains equivalent. As we continue this procedure the ifrared fixed point in the electric theory moves to strong coupling while the infrared fixed point in the magnetic theory moves to weak coupling. When the number of flavors in the electric theory becomes $F\le 3N/2$, the conformal fixed point disappears. The magnetic description now has $F\ge 3\tilde N$ and asymptotic freedom is lost. Nevertheless the infrared duality still holds as long as $N+2\le F\le 3N/2$. 

\section{Supersymmetry breaking}
\label{sec:DSB}
We now turn our attention to theories with spontaneous supersymmetry breaking. 
We will be especially interested in models where supersymmetry is broken dynamically, {\em i.e.} theories where at the classical level potential posseses supersymmetric ground states,  yet dynamical quantum effects  modify the potential and the full quantum theory either does not have any supersymmetric ground states or at least has long lived local minima with spontaneously broken supersymmetry. Due to the non-renormalization theorems discussed earlier, the existence of a SUSY vacuum at the classical level implies its existence to all orders in the perturbation theory and dynamical supersymmetry breaking (DSB) is always a non-perturbative effect.

\subsection{\oraf model}
\label{sec:oraf}
The simplest example \cite{O'Raifeartaigh:1975pr} of spontaneous supersymmetry breaking  in an interacting theory is a model with  three chiral superfields and the superpotential
\begin{equation}
\label{eq:ORW}
 W=X\left(\frac{\lambda}{2}\Phi^2-\mu^2\right)+m\Phi Y\,.
\end{equation}
Note that the model possesses an $R$-symmetry under which fields carry the following charges
\begin{equation}
 R_X=2,~~~R_Y=2,~~~R_\Phi=0\,.
\end{equation}

The $F$-term equations are
\begin{equation}
\begin{split}
& \frac{\pd W}{\pd X}=\left(\frac{\lambda}{2}\Phi^2-\mu^2\right)=0\,,\\
&\frac{\pd W}{\pd Y}=m\Phi=0\,,\\
&\frac{\pd W}{\pd \Phi}=\lambda\Phi X+mY=0\,.
\end{split}
\label{eq:orafeom}
\end{equation}
The first two of these equations are incompatible and the scalar potential of the theory has no supersymmetric ground state:
\begin{equation}
V=\left|\frac{\pd W}{\pd X}\right|^2+\left|\frac{\pd W}{\pd Y}\right|^2+\left|\frac{\pd W}{\pd \Phi}\right|^2>0\,.
\end{equation}
On the other hand, the last equation in (\ref{eq:orafeom}) always has a solution, $X=-mY/(\lambda\Phi)$,
 leaving the vev of $X$ arbitrary. It is a flat direction of the tree level potential, however, it is quite different from the moduli space of supersymmetric vacua. 
As we shall see shortly,  this flat direction is lifted in perturbation theory due to SUSY breaking.
 We will therefore refer to $X$ as a pseudo-modulus.

Let us first analyze the properties of the ground states of the theory. 
To simplify the analysis, let's assume that
 $m$ is large so that the extrema of the classical potential are found at $\Phi=Y=0$ with an arbitrary $X$. This implies that $F_\Phi=F_Y=0$ and $F_X=\mu^2$. 
The spectrum of states in the theory depends on the vev of $X$:
\begin{equation}
\begin{aligned}
 \text{scalars:~~~} &0, 0, \frac{1}{2}\left(2m^2+\lambda^2 |X|^2-\lambda \mu^2+ D(-1)\right), \frac{1}{2}\left(2m^2+\lambda^2|X|^2+\lambda\mu^2-D(1)\right),\\ &\frac{1}{2}\left(2 m^2+\lambda^2|X|^2-\lambda\mu^2-D(-1)\right), \frac{1}{2}\left(2m^2+\lambda^2|X|^2+\lambda\mu^2+D(1)\right)\,,\\
 \text{fermions:~~}&0,\frac{1}{2}\left(2m^2+\lambda^2|X|^2+D(0)\right),\frac{1}{2}\left(2m^2+\lambda^2|X|^2-D(0)\right)\,,
\end{aligned}
\end{equation}
where $D(s)=\sqrt{4\lambda^2 m^2|X|^2+(\lambda^2|X|^2+s \lambda \mu^2)^2}$.
It is easy to see that 
\begin{equation}
 \Tr[M^2_\text{scalars}]=\Tr[M^2_\text{fermions}]\,.
\end{equation}
which can be expressed in terms of a {\em supertrace}
\begin{equation}
\label{eq:orafstr}
 \Str M^2=\Tr(-1)^F M^2=0.
\end{equation}
In fact the supertrace condition is a general tree-level property of theories with spontaneous supersymmetry breaking.
 
Our next step is to calculate leading perturbative contributions to the $X$ mass by calculating one loop corrections to the vacuum energy.  A quartically divergent contribution vanishes since the theory has an equal number of bosonic and fermionic states. The quadratically divergent contribution vanishes due to the vanishing supertrace (\ref{eq:orafstr}). We then have
\begin{equation}
V_\text{eff}= \frac{1}{64\pi^2}\Str \mathcal{M}^4\log \frac{\mathcal{M}^2}{\Lambda^2} \,.
\end{equation}
It is easy to verify that while $\Str M^4$ is non-vanishing it is finite and $X$-independent.
Therefore, the logarithmically divergent contribution to the potential also vanishes. We are left with a finite one loop correction to the pseudo-modulus potential and see that it acquires a positive mass squared
\begin{equation}
 V(X)=\frac{\lambda^2}{48\pi^2}\frac{\mu^4}{m^2}|X|^2+\Order{|X|^4}\,.
\end{equation}
The vacuum is found at the origin of the field space and as a result the $R$-symmetry remains unbroken.
As was recently shown in \cite{Shih:2007av} an unbroken $R$-symmetry is a general property of \oraf models as long as all the fields in the theory have $R$-charges $0$ and $2$.

We now  consider generalizations of the \oraf model that will be of interest later in these lectures. As a first step, let us consider a theory with a global $SU(F)$ symmetry and chiral superfields transforming according to
\begin{equation}
\label{eq:oraffnplusone}
\begin{array}{|c|ccc|}
\hline
&SU(F)&U(1)_B&U(1)_R\\ \hline
B&F&1&0\\
\bar B&\bar F&-1&0\\
M&\mathrm{Adj}+1&0&2\\\hline
\end{array}
\end{equation}
For reasons that will become clear later we will refer to these fields as baryon, anti-baryon and meson. We can write the following superpotential consistent with the global symmetry
\begin{equation}
\label{eq:oraf2}
 W=\lambda \bar B_{i} M_{i j} B_j+f^2\Tr M\,.
\end{equation}
The $F$-term equations for the meson fields have the form 
\begin{equation}
\left\{ \begin{aligned}
  &\lambda \bar B_{i}B_j +f^2\delta_{ij}=0&~~~~i=j\\
  &\lambda \bar B_{ i}B_j=0&i\ne j\,.
\end{aligned}
\right.
\end{equation}
It is easy to verify directly that these equations do not have a solution and therefore SUSY must be broken. One often refers to this as a supersymmetry breaking by rank condition.
Indeed, by performing a global symmetry transformation we can guarantee that only one component of $B$, say $B_1$, has a vev. We see that the matrix $\bar B_{ i}B_j$ has rank one while rank $F$ would be needed to cancel all the $F$-terms arising from the linear term in the superpotential.

A further generalization involves a model with an $SU(\tilde N)\times SU(F)$ global symmetry\footnote{This notation anticipates the use of Seiberg duality in a future analysis of this model with gauged flavor symmetry.} and the following matter content
\begin{equation}
\renewcommand{\arraystretch}{1.3}
\label{eq:oraflast}
\begin{array}{|c|c|cc|}
\hline
&SU(\widetilde N)&SU(F)&U(1)_R\\ \hline
q&N&F&0\\
\bar q&\bar N &\bar F&0\\
M&1&\mathrm{Adj}+1&2\\\hline
\end{array}
\end{equation}
The most general renormalizable superpotential consistent with the symmetries is given by
\begin{equation}
\label{eq:oraflastW}
 W=\lambda \bar q^i_a M_i^j q^a_j +f^2 \Tr M\,,
\end{equation}
where $a$ and $i$ are $SU(\tilde N)$ and $SU(F)$ indices respectively.
We can use the global symmetry to rotate vevs of $q$ so that $\vev{q^i_a}=v_i\delta^i_a$ for $i\le \tilde min(F,\tilde N)$ and $\vev{q^i_a}=0$ otherwise.
This implies that the rank of the $\bar q q$ matrix can not exceed $min(F,\tilde N)$ while there still are $F$ non-trivial contributions to the $F$-terms arising from the linear term in the superpotential. We see that for $\tilde N<F$ supersymmetry must be broken.

The analysis of the spectrum in models with rank condition supersymmetry breaking is more involved but is similar to that in a simple \oraf model. In addition to several pseudo-moduli analogous to the  field $X$ of a simple \oraf model there exist true Goldstone bosons arising from the spontaneous breakdown of global symmetries. Nevertheless the same conclusion holds --- all pseudo-moduli obtain positive mass squareds and there exist a stable non-supersymmetric ground state in the theory \cite{Intriligator:2006dd}.
In particular the mass of $\Tr\,M$ is
\begin{equation}
m_{\Tr\,M}^2=\frac{\log 4 -1}{8\pi^2}N|\lambda^2f^2|\,,
\end{equation}
in the ground state $\Tr\,M=0$, and the $R$-symmetry remains unbroken.

\subsection{Dynamical supersymmetry breaking}
\oraf models of SUSY breaking can be used to construct phenomenological solutions of the technical hierarchy problem. However, they can not explain the origin of the hierarchy. This is because the vacuum energy in such models is an input parameter in the supersymmetric Lagrangian. On the other hand, if the supersymmetry breaking scale were determined by an energy scale associated with non-perturbative dynamics, it could be naturally small.
Thus models with dynamical supersymmetry breaking are of great interest. 

There are several guidelines in the search for dynamical SUSY breaking. The most important criterion is the value of the Witten index \cite{Witten:1982df} given by the difference between the number of bosonic and fermionic states in a theory:
\begin{equation}
 \Tr(-1)^F\equiv n_B^0-n_F^0\,.
\end{equation}
In fact, supersymmetry guarantees that the numbers of fermionic and bosonic states with non-zero energy are equal. Therefore, the value of the Witten index is determined purely by zero energy states in the theory. Moreover, the Witten index is a topological invariant of the theory. Once it is calculated for some choice of the parameters (for example in a weakly coupled regime)  the result is valid quite generally. For example, varying parameters of the theory may lift some of the ground states --- but only an equal number of fermionic and bosonic states. This has important consequences for the analysis of supersymmetry breaking.
If the Witten index is non-zero, then there exists at least one zero-energy state and supersymmetry is unbroken. On the other hand, if the Witten index vanishes, there may either be no zero-energy states or their number is even. In the former case, supersymmetry must be broken. Witten calculated the value of the index in several theories and found that it is non-zero in a pure super Yang-Mills. Therefore, pure SYM theory does not break SUSY dynamically. Furthermore, in non-chiral theories one can take all masses to be large so that low energy physics is described by super Yang-Mills. This leads us to a conclusion that non-chiral theories in general do not break SUSY. The Witten index, however, may change if the asymptotic behavior of the potential changes as some of the parameters are taken to zero or infinity. While the examples are rare, it is indeed possible for SUSY to be broken in a vector-like theory \cite{Intriligator:1996pu}.

The next step in the model-building process involves the study of classical moduli space. As we know it can only be lifted by non-perturbative dynamics. Generically, non-perturbative effects lifting the moduli space lead to runaway behavior of the scalar potential (we will discuss important counterexamples later).
Thus the most promising candidates for dynamical supersymmetry breaking are represented by  models without classical flat directions: if the moduli space at infinity is lifted by a tree level term in the scalar potential, the interplay between the tree level and non-perturbative effects may lead to SUSY breaking.

We now turn to the analysis of global symmetries. It was argued in \cite{Affleck:1983vc,Affleck:1984xz} that a theory without classical flat directions and with spontaneously broken global symmetry must break supersymmetry. To see that, recall that spontaneously broken global symmetry leads to the appearance of a Goldstone boson. Unbroken supersymmetry requires that a second real scalar field living in the same supermultiplet as the Goldstone has no potential. Changes in the vev of this scalar describe the motion along the flat direction --- contradicting our initial assumption. 

In model building, two additional conditions are often imposed --- calculability and genericity. In this context calculability means that by choice of parameters the SUSY breaking scale can be made arbitrarily small compared to the scale of strong gauge dynamics; thusthe gauge dynamics can be integrated out and the low energy physics may be described by a Wess-Zumino model. Genericity means that once all the non-perturbative effects are taken into account the superpotential is the most generic holomorphic function of the superfields consistent with symmetries of microscopic theory. In this class of models, spontaneously broken $R$-symmetry is a sufficient condition of dynamical supersymmetry breaking \cite{Nelson:1993nf}. 

We will now consider several explicit examples of dynamical supersymmetry breaking.
We begin by introducing the 3--2 model \cite{Affleck:1984xz}, probably the simplest calculable model of dynamical supersymmetry breaking. We then give an example of strongly interacting $SU(5)$ theory where supersymmetry breaking can be established by several arguments \cite{Affleck:1983vc,Meurice:1984ai} but details of the low energy physics are not calculable. Finally we discuss an Intriligator-Thomas-Izawa-Yanagida (ITIY) model \cite{Intriligator:1996pu} which breaks supersymmetry despite violating several of our guidelines.

\vskip 0.2cm
{\it $3-2$ Model of dynamical supersymmetry breaking}
 \vskip 0.2cm

Consider a theory \cite{Affleck:1984xz} with an $SU(3)\times SU(2)$ gauge group  and matter fields transforming under gauge and global symmetries according to~\footnote{It is interesting to note that $U(1)_Y$ could be gauged provided a new field, $\bar e$, with charges $(1,1,2)$ is added to the theory. This addition would not affect our discussion of supersymmetry breaking, yet, curiously enough, turn our simplest example of DSB  into a one generation version of supersymmetric Standard Model.}
\begin{equation}
\begin{array}{|c|cccc|}
 \hline
&SU(3)&SU(2)&U(1)_Y&U(1)_R\\
\hline Q&3&2&1/3&-1\\
\bar u&\bar 3&2&-4/3&-8\\
\bar d &\bar 3&1&2/3&4\\
L&1&2&-1&-3\\
\hline
\end{array}
\label{eq:threetwo}
\end{equation}
 We will also add tree level superpotential
\begin{equation}
 W_\text{tree}=\lambda Q\bar d L\,.
\end{equation}
In the limit $\Lambda_3\gg \Lambda_2$, the non-perturbative effects are captured by including the Affleck-Dine-Seiberg superpotential generated by $SU(3)$ dynamics
\begin{equation}
 W=\frac{\Lambda_3^7}{\mathrm{det} \left(Q\bar Q\right)}\,.
\end{equation}
It is easy to check that the tree level superpotential lifts all classical $D$-flat directions while presence of the non-perturbative term guarantees that the ground state is found away from the origin of the field space. 
A simple scaling argument shows that at the minimum
\begin{equation}
 Q\sim\bar u\sim\bar d\sim L\sim \frac{\Lambda}{\lambda^{1/7}},~~~~E\sim \lambda^{10/7}\Lambda^4\,.
\end{equation}
This is sufficient to conclude that the $R$-symmetry and, therefore, supersymmetry is broken. 
For small Yukawa coupling $\lambda$ the vacuum is found at large field vevs and the theory is weakly coupled and completely calculable. An explicit minimization of the potential can be performed to confirm the existence of SUSY breaking ground state and calculate the spectrum of light degrees of freedom.

In vicinity of the ground state the \kahler potential is nearly canonical in terms of elementary fields while light degrees of freedom are given by projections of these fields onto $D$-flat directions of the theory. Therefore, it is often convenient to work in terms of gauge invariant composites
\begin{equation}
 X_1=Q\bar d L,~~~X_2=Q\bar u L,~~~ Y=\mathrm{det} (\bar Q Q)\,.
\end{equation}
In terms of these variables the superpotential becomes
\begin{equation}
 W=\frac{\Lambda^7_3}{Y}+\lambda X_1\,.
\end{equation}
The \kahler potential is a bit more complicated \cite{Affleck:1984xz,Bagger:1995ay}:
\begin{equation}
K=24\frac{A+Bx}{x^2}\,,
\end{equation}
where 
\begin{equation}
\begin{aligned}
A&=\frac{1}{2}(X_1^\dagger X+X_2^\dagger X_2)\,,\\
B&=\frac{1}{3}\sqrt{Y^\dagger Y}\,,\\
x&\equiv 4\sqrt{B}\cos\left(\frac{1}{3}\arccos \frac{A}{B^{3/2}}\right)\,.
\end{aligned}
\end{equation}
We see that in low energy effective field theory the supersymmetry breaking is described in terms of a simple, albeit somewhat unusual, \oraf model with negative powers of fields in the superpotential.

\vskip 0.2cm
{\it DSB in strongly interacting models}
\vskip 0.2cm
It is possible to find strongly interacting gauge theories which satisfy our general guidelines for supersymmetry breaking. As an example, consider an $SU(5)$ theory with one superfield in $10$ and one in $\bar 5$ representation of the gauge group \cite{Affleck:1983vc,Meurice:1984ai}. The theory also possesses global $U(1)\times U(1)_R$ symmetry under which fields carry charges $(1,1)$ and $(-3,-9)$ respectively.
No classical superpotential can be written down but the $D$-term potential does not have flat directions. Both of these conclusions follow from the fact that one can not form gauge invariant operators out of single $10$ and $\bar 5$. If a supersymmetric ground state exist, it must be located near the origin of the field space where both global symmetries are unbroken.

On the other hand, it is expected that the theory confines and at low energies the physics is described by gauge invariant degrees of freedom. The consistency of the theory requires that these light composites  reproduce triangle anomalies of microscopic physics. In \cite{Affleck:1983vc} it was shown that the anomaly matching conditions require a rather large set of massless fermions: at least five if charges are required to be less than 50. This makes it quite implausible that the full global symmetry remains unbroken. But if the global symmetry is broken, so is supersymmetry. 

An independent argument for supersymmetry breaking may be obtained by deforming the $SU(5)$ model \cite{Murayama:1995ng}. Specifically, one can add an extra generation of fields in $5$ and $\bar 5$ representations. 
In the perturbed theory with the most general renormalizable superpotential (incuding a small mass $m$ for additional fields), one can show that supersymmetry is broken. One can then take the limit $m\rightarrow \infty$ and arrive at the original $SU(5)$ model. Assuming the absence of a phase transition as a function of mass one concludes that SUSY is broken.

\vskip 0.2cm
{\it DSB on quantum modified moduli space}
\vskip 0.2cm

In our last example of dynamical supersymmetry breaking we will discuss ITIY model \cite{Intriligator:1996pu} which illustrates the possibility of DSB in non-chiral theories as well as in theories with classical flat directions.
Consider an $SU(2)$ gauge theory with 4 gauge doublet and 6 gauge singlet superfields, $Q_i$ and $S_{ij}$ respectively. We will assume that singlets $S$ transform in an antisymmetric representation of $SU(4)_F$ flavor symmetry. 
We will write down a classical superpotential in terms of elementary degrees of freedom 
\begin{equation}
\label{eq:itiytree}
 W=\lambda S_{ij}Q_iQ_j\,.
\end{equation}
This superpotential lifts all $D$-flat directions of the $SU(2)$ gauge theory, however flat directions associated with singlets $S$ remain. Thus the model does not satisfy our guidelines for dynamical supersymmetry breaking: it is non-chiral and it has classical flat directions. 

At the non-perturbative level the model possesses a quantum modified moduli space. In an $SU(2)$ gauge theory quantum modified constraint (\ref{eq:feqnconstraint}) can be implemented with the following superpotential
\begin{equation}
 \label{eq:sutwoconstraint}
 W=A(\mathrm{Pf} M - \Lambda^4)\,,
\end{equation}
where $M_{ij}$ represent the 6 mesons that can be formed out of four gauge doublets, and $\Lambda$ is a dynamical scale of microscopic theory. Writing down the classical superpotential (\ref{eq:itiytree}) in terms of mesons $M$ we can see that the low energy physics is described by an \oraf model of supersymmetry breaking:
\begin{equation}
 W=A(\Pf M -\Lambda^4)+\lambda S_{ij}M_{ij}\,.
\end{equation}
As usual, there is a flat direction and we need to verify that there is no runaway behaviour as $S\rightarrow\infty$.
To do so we consider non-perturbative dynamics at a generic point on the moduli space. Let's assume that singlets $S$ obtain  large vevs giving mass to all quark superfields. We can integrate out heavy superfields and describe the low energy physics in terms of pure super Yang Mills theory with the dynamical scale
\begin{equation}
 \Lambda_{L}^6=\lambda \mathrm{Pf}S\,\Lambda^4\,,
\end{equation}
where $\Lambda_L$ is a dynamical scale of low energy SYM theory. Gaugino condensation in low energy effective theory generates the superpotential
\begin{equation}
 W=\Lambda_L^2=\lambda \left(\mathrm{Pf} S\right)^{1/2} \Lambda^2\,.
\end{equation}
It is easy to verify that the scalar potential is independent of the modulus field $\tilde S =(\mathrm{Pf} S)^{1/2}$ and is non-vanishing, $V=\lambda^2 \Lambda^4$. 

Furthermore, at large $S$ corrections to the scalar potential are perturbative \cite{Shirman:1996jx} and the pseudo-flat direction is lifted
\begin{equation}
 V=\frac{\lambda^2}{Z_S}\Lambda^4=\lambda^2\Lambda^4\left(1+\frac{2}{16\pi^2}\ln\frac{|\tilde S|^2}{M_{\mathrm{UV}}}+\Order{\lambda^4}\right)\,.
\end{equation}

The analysis of the Coleman-Weinberg potential near the origin of the moduli space is more complicated since a priori the strong coupling dynamics may be important. This analysis was performed in \cite{Chacko:1998si} and it was found that uncalculable corrections due to strong dynamics are negligible and the $\tilde S$ potential is
\begin{equation}
 V=\frac{5\lambda^4\Lambda^2}{16\pi^2}(2\ln 2-1)|\tilde S|^2+\Order{S^4}\,.
\end{equation}
We conclude that the potential of the ITIY model is calculable both for small and large $\tilde S$ and a ground state is found at $S=0$. On the other hand the approximations made in the above calculations break down when $\lambda S\sim\Lambda$ leaving the possibility that another minimum of the potential exists with $S\sim\Lambda/\lambda$.

This model can be generalized to other examples with quantum modified moduli space, most straightforwardly to $SU(N)$ theories with $F=N$ flavors and $SP(2N)$ theories with $N+1$ flavors.

\subsection{Metastable SUSY breaking}
Our discussion of dynamical supersymmetry breaking makes it clear that this is not a generic effect in SUSY gauge theories. Moreover, once the DSB sector is coupled to SUSY extensions of the Standard Model, one typically finds that supersymmetric vacua reappear elsewhere on the field space while SUSY breaking minima survive only as metastable, if long-lived, vacua. 

If metastability is inevitable, then it is reasonable to accept it from the start. Indeed it was shown recently by Intriligator, Seiberg, and Shih that metastable minima of the potential with broken SUSY are quite generic \cite{Intriligator:2006dd} and often arise in very simple models. 
Probably the simplest example is SUSY QCD with $N$ colors and $F=N+1$ massive flavors, a theory we already discussed in section \ref{sec:SUSYQCD}. 
In the presence of the mass term the full superpotential of the model, including effects of the non-perturbative dynamics is
\begin{equation}
\label{eq:ISS}
 W=\frac{1}{\Lambda^{2N-1}}\left(B M\bar B-\det M\right)+ m\Tr M\,,
\end{equation}
where $m$ is the quark mass term, $\Lambda$ is the strong coupling scale of the theory while $M$, $B$, and $\bar B$ are mesons (\ref{eq:meson}) and baryons (\ref{eq:baryon}).
Near the origin of the field space baryons and mesons are weakly coupled degrees of freedom with canonical kinetic terms. This means that we can read off dimensions of the operators directly from the superpotential (\ref{eq:ISS}). We see that $B M\bar B$ term in the superpotential is a marginal operator in the infrared while the quark mass term is a relevant operator, $m\Lambda \Tr\, M$, in terms of canonically normalized meson field $M$. On the other hand, $\det M$ remains irrelevant (except in the case of $N=2$) and can be neglected in the discussion of  dynamics near the origin of the moduli space.
It is now easy to notice that the low energy dynamics of this theory is well described by an \oraf model (\ref{eq:oraffnplusone}) with an identification $f^2=m\Lambda$. 
Near the origin of field space and as long as mass term $m$ is sufficiently small effects of strong gauge dynamics are negligible compared to terms in Coleman-Weinber potential and the analysis of section \ref{sec:oraf} remains valid.
On the other hand, in the UV (or at large field vevs) the theory deconfines and presence of the $\det M$ in the superpotential leads to restoration of supersymmetry at
\begin{equation}
 \vev{M}=\left(m\Lambda^{2N-1}\right)^{1/N}\unit_{F}\,.
\end{equation}
The effects of the strong gauge dynamics become important when $m \sim \Lambda$ and it is reasonable to restrict ourselves to small masses $m\ll \Lambda$. In this case $\vev{M}\ll \Lambda$ and the perturbative calculations are reliable not only near the origin of the fields space but also in the vicinity of supersymmetric ground states.

Following \cite{Intriligator:2006dd} we can generalize this example by  gauging the global $SU(\widetilde N)$ symmetry of the model defined by (\ref{eq:oraflast}) and (\ref{eq:oraflastW}). We will also choose $F\ge 3\widetilde N$ and identify this model with the magnetic description of an asymptotically free $SU(N)$ theory. The fields of our model, $M$, $q$, and $\bar q$, are then mesons, as well as magnetic quarks and antiquarks. The first term in (\ref{eq:oraflastW}) is generated by the strong dynamics in the electric theory while the second term corresponds to the quark mass in the electric description, $f^2\Tr\,M\sim m \Tr\, Q\bar Q$.
To verify that the analysis of section \ref{sec:oraf} remains valid in the presence of newly introduced gauge dynamics, we note that for our choice of $F$ and $\widetilde N$ the magnetic description is IR free, gauge dynamics of the $SU(\widetilde N)$ theory is weakly coupled near the origin of the field space and we are still justified in performing perturbative calculation. We must include contributions of gauge supermultiplet in our calculation of Coleman-Weinberg potential, however, to leading order in SUSY breaking parameter it has a supersymmetric spectrum and our earlier results are not modified.

On the other hand, from the analysis of electric description we know that supersymmetric vacua exist in this theory. They can also be found in magnetic description by carefully examining the effects of gauge dynamics at large field vevs.
Indeed, at large $M$ magnetic quarks become massive and can be integrated out, leading to gaugino condensation in magnetic theory and the effective superpotential, $W=\Lambda_L^3$, where $\Lambda_L$ is a strong coupling scale of a low energy $SU(\widetilde N)$ SYM theory. Using the fact that holomorphic gauge coupling evolves only at one loop, this superpotential can be written as
\begin{equation}
 W=\Lambda_L^3=\left(\mu^{3\widetilde N}e^{-8\pi^2/g_L^2(\mu)}\right)^{1/\widetilde N}=\left(
\Lambda_{UV}^{3\widetilde N-F} \det M  e^{-8\pi^2/\tilde g^2(\Lambda_{UV})}\right)^{1/\widetilde N}=\frac{\det M}{\widetilde \Lambda^{F-3\widetilde N}}\,,
\end{equation}
where $\widetilde \Lambda$ is the scale of the Landau pole in the magnetic theory.
Together with the superpotential (\ref{eq:oraflastW}) this dynamical term leads to restoration of supersymmetry.
Just as in our previous example, for sufficiently small mass terms in electric theory, the potential  is fully calculable both near supersymmetric and supersymmetry breaking minima.

It is important for phenomenological applications that the metastable non-supersymmetric vacua are sufficiently long-lived. The semi-classical decay probability of a false vacuum is given \cite{Coleman:1977py} by $\exp(-S)$, where $S$ is a bounce action.
For the models discussed in this section the bounce action was estimated in \cite{Intriligator:2006dd}:
\begin{equation}
 S\sim\left(\frac{\Lambda}{m}\right)^{2(F-N)/(F-N)}\gg 1\,.
\end{equation}
This suggests that generically it is possible to achieve sufficiently long lifetime of the vacuum by appropriate choice of the parameters. One still needs to explain why the non-supersymmetric ground state is chosen in the early Universe. It was argued in \cite{Abel:2006cr}  that   a non-supersymmetric vacuum is generically preferred over a supersymmetric one in the early Universe due to thermal effects. This conclusion holds as long as a metastable vacuum is closer to the origin of the field space than a supersymmetric one.

\subsection{Fayet-Illiopolous model}
For completeness, we will briefly introduce an example of spontaneous SUSY breaking with a non-vanishing D-term \cite{Fayet:1974jb}.
Consider a $U(1)$ theory with the Lagrangian
\begin{equation}
\begin{split}
 \Lagr=&\int d^4\theta \Phi_+^\dagger e^V\Phi_++\Phi_-^\dagger e^{-V}\Phi+(\int d^2\theta \frac{1}{4g^2}\Wg^\alpha \Wg_\alpha+\hc)\\
&+\int d^4\theta \xi^2 V+(\int d^2\theta m\Phi_+\Phi_-+\hc)\,.
\end{split}
\end{equation}
In this model the D-term equation of motion
\begin{equation}
D=g(\left|\Phi_+\right|^2-\left|\Phi_-\right|^2+\xi^2)=0
\end{equation}
requires non-zero vev for $\Phi_-$. On the other hand thr F-term conditions
\begin{equation}
 F_{\Phi_+}=m\Phi_-=0,~~~F_{\Phi_-}=m\Phi_+=0
\end{equation}
require that both fields vanish. Thus the potential at the minimum is non-vanishing and SUSY is broken. It is straightforward to verify that the tree level spectrum of the theory satisfies the supertrace condition, $\Str\mathcal{M}^2=0$.

\subsection{Goldstino}
According to the Goldstone theorem spontaneously broken symmetries must always lead to an appearance of massless particles with derivative interactions. In theories of spantaneously broken supersymmetry the broken generator is a spinor and the corresponding massless particle is a fermion, goldstino.
Indeed, presence of massless goldstino is required by the goldstino theorem which can be derived quite easily for a general supersymmetric model \cite{Salam:1974zb,Witten:1981nf}.

The goldstino is always a fermion in the supermultiplet that leads to SUSY breaking, in other words it is a fermion in a supermultiplet with a non-vanishing $F$- or $D$-term. If several supermultiplets acquire $F$- and $D$-term vevs in the ground state of the theory, goldstino is a linear combination of fermions living in these supermultiplets.
To illustrate the appearance of a goldstino, consider a theory with gauginos $\lambda^a$,  and matter fermions $\chi_i$. In all examples of SUSY breaking we have considered so far, fermion mass matrix is not affected by supersymmetry breaking to leading order in SUSY breaking parameters $F$ and $D$. It has the form
\begin{equation}
\label{eq:goldstinomassmatrix}
\left(\begin{array}{cc}
  0&\sqrt{2} g_a \varphi_i^*\lambda^a T^a_{ij}\chi_j\\
  g_a \varphi_i^*\lambda^a T^a_{ij}\chi_j& W_{ij}\\   
               \end{array}\right)\,,
 \end{equation}
and off-diagonal termes appear whenever gauge symmetry is broken by vev of $\varphi$. 
When SUSY is broken this mass matrix has at least one zero eigenvalue whose eigenvector is given by 
\begin{equation}
 \left(\begin{array}{c}{D^a}/{\sqrt{2}}\\
        F_i
       \end{array}
 \right)\,.
\end{equation}
This eigenvector is non-trivial whenever at least one $F$- or $D$-term is non-vanishing and corresponds to the goldstino which can be written as
\begin{equation}
 \goldstino=\frac{1}{F_{\goldstino}}\left(\frac{D^a}{\sqrt{2}}\lambda^a+F_i\chi_i\right)\,,
\end{equation}
where
\begin{equation}
 F_{\goldstino}=\sqrt{\sum_a\frac{(D^a)^2}{2}+\sum_i |F_i|^2}\,.
\end{equation}

The goldstino effective Lagrangian can be written as
\begin{equation}
 \Lagr_\text{goldstino}
=i\goldstino^\dagger\pd_\mu\sigmabar^\mu\goldstino+\frac{1}{F_\goldstino}\goldstino^\alpha \pd_\mu j^\mu_\alpha\,,
\end{equation}
where the supercurrent $j^\mu_\alpha$ is 
\begin{equation}
 j^\mu_\alpha=(\sigma^\nu\sigmabar^\mu \eta_i)_\alpha D_\nu \varphi^{*i}-\frac{1}{2\sqrt{2}}(\sigma^\nu\sigmabar^\mu\sigma^\rho\lambda^{\dagger a})_\alpha F^a_{\nu\rho}\,.
\end{equation}

Finally, we would like to mention the role of a goldstino in locally supersymmetric theories.
 As we know such theories require the existence of a gravitino, a spin 3/2 superpartner of the graviton. Once supersymmetry is broken, the gravitino becomes massive by eating the goldstino, in a complete analogy to Higgs mechanism where the gauge boson becomes massive by eating a Goldstone boson.

\section{Minimal Supersymmetric Standard Model}
\label{sec:MSSM}
\subsection{Matter content and interactions}
We will now study a  Minimal Supersymmetric Standard Model (MSSM). Supersymmetry requires that all the Standard Model particles are accompanied by superpartners. In the gauge sector we must include gauginos in the adjoint representation for each of the Standard Model gauge groups. The matter content is given by chiral superfields with the following charge assignments 
\begin{equation}
\label{eq:MSSMmatter}
 \begin{array}{|c|ccr|}
  \hline
 & SU(3)&SU(2)&U(1)_Y\\
\hline
Q_f&3&2&1/3~\\
\bar u_f&\bar 3&1&-4/3~\\
\bar d_f&\bar 3 &1&2/3~\\
L_f&1&2&-1~\\
\bar E_f&1&1&2~\\
H_u&1&2&1~\\
H_d&1&2&-1~\\
\hline
 \end{array}
\end{equation}
where $f=1..3$ is a generation index.
With a little abuse of notation we will use the same symbol both for the superfields and Standard Model particles (i.e. fermions and Higgses) while denoting the superpartners (sfermions and Higgs superpartners, higgsinos) with a tilde.  
An important feature of matter content (\ref{eq:MSSMmatter}) is the presence of two Higgs multiplets. There are two reasons for this. First of all, higgsinos carry $SU(2)\times U(1)_Y$ quantum numbers and contribute to anomalies. Since the SM is anomaly free in the absence of higgsinos, two higgsinos with opposite charges represents a minimal extension  without new contributions to anomalies. Furthermore, two Higgs supermultiplets are required to reproduce all the SM Yukawa couplings. 

Indeed, in the Standard Model, due to the fact that $\bar 2$ and $2$ representations of the $SU(2)$ gauge group are equivalent, a single Higgs boson is sufficient to write down both the up and down type Yukawa matrices. In MSSM, however, Yukawas must arise from holomorphic terms in the superpotential and complex conjugate of $2$ representation can not appear in the superpotential. Thus full set of Yukawa couplings is only possible in the presence of two Higgs doublets and  is contained in the following superpotential
\begin{equation}
\label{eq:MSSMYukawa}
 W_\text{Yukawa}=\lambda^u_{f\fprime} H_u Q_f \bar u_\fprime+\lambda^d_{f\fprime}H_dQ_f\bar d_\fprime+\lambda^L_{f\fprime}H_u L_f\bar e_\fprime \,.
\end{equation}
For example, the top Yukawa coupling, $\lambda_t H_u  Q_3 \bar t$, is contained in the first term while the bottom Yukawa, $\lambda_b H_d Q_3 \bar b$, is contained in the second. 

In addition to the superpotential of eqn. (\ref{eq:MSSMYukawa}), two more types of terms are allowed by the symmetries. First, we can write a supersymmetric Higgs mass term
\begin{equation}
\label{eq:Wmu}
 W_H=\mu H_u H_d\,.
\end{equation}
As we will see shortly this term  is important for generating electroweak symmetry breaking but at the same time it leads to a well-known $\mu$-problem.

\begin{figure}
\begin{center}
 \includegraphics[width=2.5in]{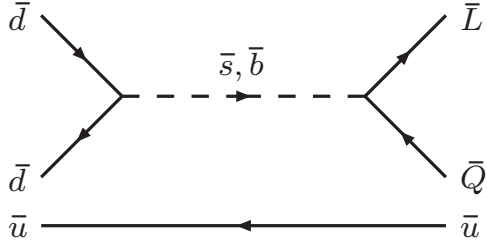}
\end{center}
\caption{Proton decay diagram arising from R-symmetry violating interactions}
\label{figprotondecay}
\end{figure}

Another set of allowed superpotential interactions: 
\begin{equation}
 W_\Rsb=\alpha^{ijl}Q_iL_j\bar d_k+\beta^{ijk}L_iL_j\bar E_k+\gamma^iL_iH_u+\delta^{ijk}\bar d_i \bar d_j\bar u_k\,,
\end{equation}
where $\beta$ and $\delta$ are antisymmetric under the interchange $(i\leftrightarrow j)$ is much more dangerous.
Terms in $W_\Rsb$ are renormalizable and violate both lepton and baryon numbers. They immediately lead to proton decay through the diagram depicted in Figure \ref{figprotondecay}. A rough estimate of the proton lifetime gives
\begin{equation}
 \begin{split}
  \Gamma_p\sim &\frac{|\alpha \delta|^2}{8\pi^2}\frac{m_p^5}{m_{\tilde q}^4}\,,\\
\tau_p=&\frac{1}{\Gamma_p}\sim\frac{1}{|\alpha\delta|^2}\left(\frac{m_{\tilde q}}
{\text{1 TeV}}\right)^4 2\times 10^{-11}s\,,
 \end{split}
\end{equation}
where $m_{\tilde q}$ is a squark mass. Comparing this result with experimental limits on proton lifetime we see that either coupling constants in $W_\Rsb$ must be extremely suppressed, $|\alpha\delta|<10^{-25}$, or the SUSY breaking scale (parameterized here by $m_{\tilde q}$) is very large, $m_{\tilde q}^2> 10^{31}\, \GeV^2$.

One usually approaches this problem is by introducing a new discrete symmetry which forbids dangerous couplings. Such a symmetry can be thought of as a discrete subgroup of an R-symmetry, called R-parity. Under R-parity all the Standard Model particles (including both Higgs boson doublets) are even while all the superpartners are odd. Interestingly, $R$-parity can be introduced without reference to $R$-symmetry by defining charges of particles under $R$ parity as a combination of their  fermion number and $B-L$ charge
\begin{equation}
 R=(-1)^{(B-L)+F}\,.
\end{equation}

In addition to suppressing proton decay, R-parity leads to several important consequences:
\begin{itemize}
 \item At colliders superpartners are produced in pairs;
 \item The lightest superpartner is stable and if it happens to be neutral, provides an excellent dark matter candidate;
 \item Each sparticle other than LSP will decay to an odd number of LSP's (plus ordinary particles).
\end{itemize}

We would like to comment that one can consider R-parity violating extensions of the SM as long as dangerous couplings are fine-tuned to be small. While this class of models may lead to interesting experimental signatures its is beyond the scope of these lectures.

\subsection{Soft SUSY breaking}
\label{sec:softbreak}
We now introduce SUSY breaking into the MSSM. The supersymmetry breaking must be {\em soft}, that is it should not reintroduce quadratic divergencies. This could be achieved if SUSY breaking is a spontaneous symmetry breaking. One could attempt to construct extensions of the MSSM whith spontaneously broken supersymmetry. As we know tree level spectrum of such models satisfies the supertrace condition, $\Str\, \mathcal{M}=0$. While the supertrace condition is modified by quantum effects, within MSSM alone such modifications are small since the Standard Model is a weakly interacting theory at EWSB scale.
 Since the Standard Model fermions are generally light, the supertrace condition requires the existence of new light bosons which have not been observed experimentally \cite{Dimopoulos:1981zb}. We conclude that SUSY must be broken in a different, {\em hidden}, sector of the theory.

\begin{figure}[t]
 \begin{center}
  \includegraphics{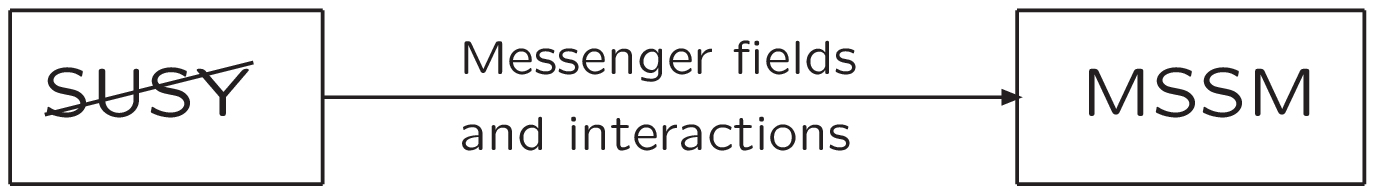}
 \end{center}
\caption{Typical structure of supersymmetric extensions of the Standard Model}
\label{figsusybreakingdiagram}
  \end{figure}

We will consider a scenario of SUSY breaking depicted in Fig. \ref{figsusybreakingdiagram}. We will imagine that supersymmetry is broken in a hidden sector by one of the mechanisms described in section \ref{sec:DSB}. We will then have to introduce interactions between the hidden and visible sectors that will communicate SUSY breaking to the MSSM fields and generate superpartner masses. While there are several different mechanisms that could mediate SUSY breaking to the Standard Model sector, a general form of soft-breaking terms can be obtained from the following argument.
 Imagine integrating out hidden sector physics completely and obtaining an effective Lagrangian in MSSM sector. In such a Lagrangian coupling constants become functions of integrated out hidden sector superfields. 
We can now use a familiar trick of promoting Lagrangian parameters to background superfields. To include the effects of spontaneous SUSY breaking in the hidden sector, we will simply allow these background superfields to have non-vanishing $F$-terms\footnote{Hidden sector $D$-terms may also be considered.}.
As an example consider (\ref{eq:Wmu}). Promoting $\mu$ to a superfield $\mu\rightarrow \mu +B\theta^2$ leads to the soft SUSY breaking terms in the Lagrangian
\begin{equation}
\Lagr_B=B h_uh_d+\hc\,. 
\end{equation}
This term represents the new SUSY breaking  masses in the Higgs potential and together with the $\mu$-term plays an important role in electroweak symmetry breaking.
More generally, we can promote all the Yukawa and gauge couplings to functions of background superfields
\begin{equation}
\label{eq:yukawaspurion}
  \lambda_{ff^\prime}^u\rightarrow \lambda_{ff^\prime}^u(\Sigma/M),
~~~\lambda^d_{ff^\prime}\rightarrow \lambda^d_{ff^\prime}(\Sigma/M),
~~~\lambda^L_{ff^\prime}\rightarrow \lambda^L_{ff^\prime}(\Sigma/M),
~~~\frac{1}{g_a^2}\rightarrow \frac{1}{g_a^2\left(\Sigma/M\right)}\,,
\end{equation}
where $M$ is a characteristic scale of interactions between the Standard Model and SUSY breaking sector and $\Sigma$ represents hidden sector superfields. In the simplest case, $\Sigma$ is a single superfield and the flavor structure of soft parameters is completely encoded in Yukawa matrices $\lambda_{ff^\prime}$. More generally, the Yukawa matrices and gauge coupling functions could depend non-trivially on several hidden sector superfields. For example one can have $\lambda^{u,d}_{ff^\prime}=\Sigma^{u,d}_{ff^\prime}/M$ and $1/g_a^2=\Sigma^a/M$.
In this case, suppressing flavor indices and writing $\vev{\Sigma}=\sigma+F_{\Sigma} \theta^2$ we can write Yukawa couplings as
\begin{equation}
 \lambda^u=\frac{\sigma^u}{M},~~~\lambda^d=\frac{\sigma^d}{M},~~~\lambda^L=\frac{\sigma^L}{M},~~~\frac{1}{g_a^2}=\frac{\sigma^a}{M}\,,
\end{equation}
as well as trilinear scalar interactions, so called A-terms,
\begin{equation}
 \Lagr_{A}=A^u h_u \tilde Q \tilde{\bar u}+A^dh_d\tilde Q\tilde{\bar d}+A^L h_u \tilde L\tilde{\bar e} +\hc\,,
\end{equation}
where 
\begin{equation}
\label{eq:softA}
A^u=\frac{F_{\Sigma}^u}{M},~~~A^d=\frac{F_{\Sigma}^d}{M},~~~A^L=\frac{F_{\Sigma}^L}{M}\,.
\end{equation}
Similarly, gauge couplings and gaugino masses may be written in terms of the spurion vevs
\begin{equation}
\label{eq:softg}
 \frac{1}{g^2_a}=\frac{\sigma^a}{M},~~~M^a=\frac{F_\Sigma^a}{M}\,.
\end{equation}

We still need to generate scalar masses. Let's look at the \kahler potential. Generically non-renormalizable interactions between the hidden and visible sectors will appear in the  \kahler potential of the effective theory even if they are absent in the microscopic description. As an example \kahler potential of squark superfields may take the form
\begin{equation}
\label{eq:mediationkahler}
 K=\left(\delta_{f\fprime}+c_{f\fprime}\frac{\Sigma^\dagger\Sigma}{M^2}\right) Q^\dagger_f Q_\fprime\,.
\end{equation}
This results in soft squark masses 
\begin{equation}
\label{eq:softs}
\tilde m^2_{f\fprime}=\frac{c_{f\fprime} |F_\Sigma|^2}{M^2}\,.
\end{equation}
Similar soft masses are generated for other sfermions as well as scalar components in the Higgs multiplets.

\subsection{Higgs Sector}
\label{sec:HiggsSector}
We now turn to the question of electroweak symmetry breaking in MSSM with softly broken SUSY. The Higgs potential is given by
\begin{equation} 
 V=V_D+V_F+V_{SUSY}\,,
\end{equation}
where $V_D$ is the D-term potential
\begin{equation}
 V(H_u,H_d)=\frac{g^2+g^{\prime 2}}{8}\left(|H_u^0|^2+|H_u^+|^2-|H_d^0|^2-|H_d|^2\right)^2+\frac{g^2}{2}\left|H_u^{+}H_d^{0*}+
H_u^0H_d^{-*}\right|^2,
\end{equation}
$V_F$ is the F-term potential
\begin{equation}
 V_F=\mu^2 \left(|H_u^0|^2+|H_u^+|^2+|H_d^0|^2+|H_d^-|^2\right)\,,
\end{equation}
and $V_\text{SUSY}$ represents SUSY breaking terms in the potential
\begin{equation}
\begin{split}
 V_\text{SUSY}=&\tilde m_u^2 \left(|H_u^0|^2+|H_u^+|^2\right)+ \tilde m_d^2\left(H_d^0|^2+|H_d^-|^2\right)\\
+ &B\left(H_u^{+}H_d^{-}-H_u^0H_d^0\right)+B\left(H_u^{+*}H_d^{-*}-H_u^{0*}H_d^{0*}\right)\,,
\end{split}
\end{equation}
where $\tilde m_u^2$ and $\tilde m_d^2$ are soft Higgs masses.
Electroweak symmetry breaking requires that the Higgs potential is bounded from below and has a minimum at non-vanishing vevs. The Higgs mass matrix will satisfy these conditions if
\begin{equation}
\begin{aligned} 
&|
B|^2>(\tilde m_u^2+|\mu|^2)( \tilde m_d^2+|\mu|^2)\,,\\
& 2\mu^2+\tilde m_u^2+\tilde m_d^2>2|B|\,.
\end{aligned}
\end{equation}
Typically these conditions are not satisfied at the SUSY breaking scale. However, RG evolution modifies relations between superpartner masses and may lead to radiative electroweak symmetry breaking. The dominant effect arises from the Higgs interactions with the third generation 
\begin{equation}
 \frac{d}{dt}\left(\begin{aligned}
&\tilde m_u^2\\
&\tilde m_{\tilde t}^2\\
&\tilde m_{Q_3}^2
\end{aligned}\right)=-\lambda_t^2\left(
\begin{aligned}
 3&&3&&3\\
2&&2&&2\\
1&&1&&1\\
\end{aligned}
\right)-A_t^2\left(\begin{aligned}3\\2\\1\end{aligned}\right)\,.
\label{eq:softhiggsrg}
\end{equation}
We can see that $H_u$ receives the largest negative contribution and once its mass is driven negative electroweak symmetry is broken. 

By requiring that the parameters in the Higgs potential lead to experimentally observed $Z$ and $W$ mass we obtain relations between soft parameters which must be satisfied at the weak scale:
\begin{equation}
\begin{aligned}
 \mu^2=\frac{\tilde m_u^2-\tilde m_d^2\tan^\beta}{\tan^2\beta-1}-\frac{1}{2}M_Z^2\,,\\
B=\frac{(\tilde m_u^2+\tilde m_d^2+2\mu^2)\sin 2\beta}{2}\,,
\end{aligned}
\end{equation}
where $\tan\beta=v_u/v_d$.

From this expression we see that naturalness requires that both $\mu$ and $B$ are electroweak scale parameters. 
However, the $\mu$-term is a supersymmetric term in the Lagrangian and could take any value between EWSB and Planck scales. This leads to the so-called $\mu$-problem. Only models where the $\mu$-term arises as a result of SUSY breaking are expected to avoid fine-tuning. Even then, the absence of fine-tuning is not guaranteed. To illustrate this issue, let's replace the $\mu$-term with a vacuum expectation value of a new gauge singlet field $X$ coupled to the Higgses
\begin{equation}
\label{eq:muX}
 W_X=\lambda_X X H_u H_d\,.
\end{equation}
It is possible to construct models where $\vev{X}$ is only generated as a result of SUSY breaking. 
Even in these models the coupling constant $\lambda$ often needs to be small to guarantee the correct magnitude of the $\mu$-term.
Additionally, $\vev{F_X}$ is often generated and leads to the $B$-term.
Typically one finds $F_X\sim X^2$ and the ratio between $\mu^2$ and $B$ terms is given by
\begin{equation}
 \frac{B}{\mu^2}\sim\frac{\lambda_X F_X}{\lambda_X^2 X^2}\sim \frac{F_X}{\lambda_X X^2}\sim\frac{1}{\lambda_X}\,.
\end{equation} 
In other words, if a $\mu$-term of the correct magnitude results in a phenomenologically unacceptable $B$ term. This result holds quite generically in theories were small parameters are used to generate the soft terms from the fundamental scale of SUSY breaking.

\subsection{Flavor problem}
The most general set of $R$-parity invariant soft terms leads to a model with 105 parameters in addition to those in the Standard Model itself. One would like to find an organizing principle which reduces number of parameters and makes the model predictive. Furthermore, generic points on the MSSM parameter space are ruled out by existing experiments. The most stringent limits arise from constraints on flavor violating processes.

\begin{figure}
 \begin{center}
  \includegraphics[width=2.5in]{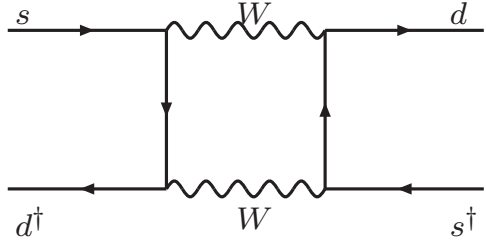}
 \end{center}
\caption{Standard Model contribution to KK mixing}
\label{figKKmixing}
\end{figure}

For example, consider $K-\bar K$ mixing. In the Standard Model, GIM mechanism ensures that the leading contribution to this process, arising through a diagram in Figure \ref{figKKmixing}, starts at order
$\Order{m_\text{quark}^2}$:
\begin{equation}
 \mathcal{M}_{K\bar K}^{SM}\approx\alpha_2^2\frac{m_c^2}{M_W^2}\sin^2\theta_c\cos^2\theta_c\,,
\end{equation}
where $\theta_c$ is the Cabbibo angle.

\begin{figure}
 \begin{center}
  \includegraphics[width=2.5in]{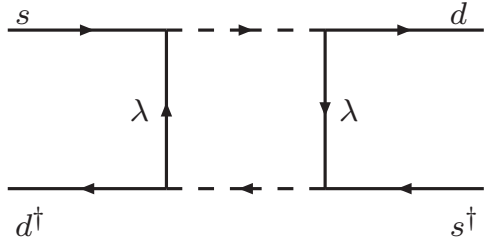}
 \end{center}
\caption{MSSM contributions to $K-\bar K$ mixing}
\label{figMSSMKK}
\end{figure}

In the MSSM additional contributions arise due to processes in Figure \ref{figMSSMKK}:
\begin{equation}
\label{eq:MSSMKK}
 \mathcal{M}^{MSSM}_{K\bar K}\approx 4\alpha_3^2\left(\frac{\Delta \tilde m_Q^2}{M^2_{SUSY}}\right)\frac{1}{M^2_{SUSY}}\,,
\end{equation}
where $M_\text{SUSY}$ is a typical scale of the soft MSSM parameters.
As we can see this contribution is formally enhanced compared to the Standard Model amplitude by a factor of the order $(\alpha_3/\alpha_2)^2$. On the other hand, any new physics contribution can't be large since the Standard Model result is consistent with experimental observations.
This implies the following relation
\begin{equation}
 \left(\frac{\Delta \tilde m^2_Q}{M_{SUSY}^2}\right)<4\times 10^{-3}\frac{M_{SUSY}}{550\GeV}\,.
\end{equation}
We conclude that the squark mass matrix must be diagonal in flavor space in the same basis as the quark mass matrix or SUSY breaking scale is much larger than electroweak scale.

\begin{figure}
 \begin{center}
  \includegraphics[width=3in]{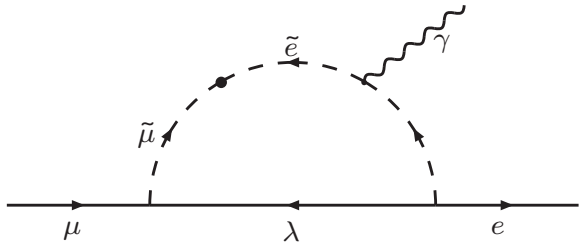}
 \end{center}
\caption{SUSY contribution to muon decay}
\label{figmuegamma}
\end{figure}

There also exist strong constraints on flavor violation in the slepton sector. In addition to the Standard Model muon decay $\mu\rightarrow e\nu\nu^*$ the generic choice of MSSM parameters introduces a new decay channel, $\mu\rightarrow e\gamma$, which proceeds through the diagram in Figure \ref{figmuegamma}. It is not difficult to estimate the branching ratio
\begin{equation}
 \frac{\Gamma_{\mu\rightarrow e\gamma}}{\Gamma_{\mu\rightarrow e\nu\nu^*}}\approx 10\times 10^{-4}\left(\frac{500\GeV}{M_{SUSY}}\right)^4\times \left(\frac{\Delta \tilde m_L^2}{M_{SUSY}^2}\right)\,.
\end{equation}
Experimentally this ratio is less than $10^{-11}$. Once again either $\Delta \tilde m^2_L$ is nearly diagonal in the same basis as the charged lepton mass matrix or the SUSY breaking scale is extremely high.

As we just indicated the flavor problem could be resolved if the fermion and sfermion mass matrices are diagonal in the same basis, in other words if a super-GIM mechanism is operational in MSSM. This can be achieved in models with flavor symmetries, see for example \cite{Nir:1993mx}. Another resolution would require that mechanism mediating SUSY breaking between hidden and visible sectors is flavor blind. We will discuss some realizations of this idea in the next section.

\section{Mediation of SUSY breaking}
\label{sec:mediation}
We finally come to the discussion of mechanisms that can communicate SUSY breaking between hidden and visible sectors. We will only discuss the three most popular mechanisms out of several interesting possibilities: supergravity, gauge, and anomaly mediation.
\subsection{SUGRA mediation}
The most minimal approach to mediating supersymmetry breaking between hidden and visible sectors is through supergravity interactions.
Generically one should expect that the most general interactions consistent with the symmetries of both hidden and visible sector will be generated in an effective theory with Planck suppressed couplings. Thus formulas of section \ref{sec:softbreak} will apply with a messenger scale $M=M_\text{Pl}$. Using $\Mpl$ in formulas   (\ref{eq:softA}), (\ref{eq:softg}), and (\ref{eq:softs}) and requiring that soft SUSY breaking parameters in the Standard Model sector are of the order TeV, implies that the fundamental SUSY breaking scale is of order $10^{11}\,\GeV$. This also leads to a gravitino mass of the order TeV.

Is the spectrum of gravity mediation consistent with FCNC constraints?  At first it appears natural to assume that all soft scalar masses are universal since gravity couples universally to all fields. Similarly, one could expect universality for A-terms as well as gaugino masses. However, this assumption is not fully justified. Indeed, one should expect that the microscopic description of fundamental theory   may contain new particles with order 1 couplings both to the hidden and visible sectors and masses of order $M_\text{Pl}$. These particles do not necessarily belong to the gravity multiplet and as such do not have to couple universally to all the Standard Model fields. Integrating out these particles leads to low energy effective description with order one flavor violations in sfermion mass matrices.

Thus one needs to impose additional conditions to guarantee the compatibility of theoretical predictions with the existing experimental observations. In the gravity mediation approach one simply assumes universality at the matching scale. Namely, one assumes universal gaugino and sfermion masses while A-terms are taken to be proportional to Yukawa matrices. The soft terms in the Lagrangian are determined by 4 parameters
\begin{equation}
\label{eq:SUGRAsoftterms}
 m_{1/2}=f \frac{F}{\Mpl},~~~m_0^2=k\frac{|F|^2}{\Mpl^2},~~~A_0=\alpha\frac{F}{\Mpl},~~~B=\beta\frac{F}{\Mpl}\,.
\end{equation}
At the matching scale sfermion masses as well as soft Higgs masses are given by
\begin{equation}
\label{eq:sugram}
 \tilde m_{Qff^\prime}^2=\tilde m_{\bar uff^\prime}^2=\tilde m_{\bar dff^\prime}^2=\tilde m_{L ff^\prime}^2=\tilde m^2_{\bar eff^\prime}=\tilde m_u^2=\tilde m_d^2=m_0^2\delta_{ff^\prime}\,.
\end{equation}
Tirlinear couplings are
\begin{equation}
\label{eq:sugraA}
 A^u_{ff^\prime}=\lambda^u_{ff^\prime}A_0,~~~A^d=\lambda^d_{ff^\prime}A_0,~~~A^L=\lambda^L_{ff^\prime}A_0\,.
\end{equation}
Finally gaugino masses are unified at the matching scale
\begin{equation}
\label{eq:sugralambda}
 M_3=M_2=M_1=m_{1/2}\,.
\end{equation}
One usually assumes that the hidden sector does not have light fields and decouples at the matching scale. As a result low energy values of soft masses are determined by the equations (\ref{eq:sugram}), (\ref{eq:sugraA}), (\ref{eq:sugralambda}) and renormalization group evolution between the matching and electroweak scales. In particular, renormalization group evolution drives $H_u$ mass squared negative according to (\ref{eq:softhiggsrg}).
However, it is possible that hidden sector is both strongly interacting and contains particles much lighter that matching scale. In this case, effect of the hidden sector RG evolution can not be neglected\cite{Cohen:2006qc}. This situation is not unique to supergravity and effects of hidden sector renormalization on soft parameters may be significant in other mediation mechanisms.

To conclude this section we briefly mention the status of $\mu$ problem in gravity mediation. It turns out that gravity mediation allows for a rather elegant solution of the  $\mu$ problem \cite{Giudice:1988yz}. First, it is quite easy to forbid appearance of a large $\mu$-term by imposing some symmetry on the Lagrangian. This can be an $R$-symmetry, PQ-symmetry or a discrete symmetry. One can then introduce Planck suppressed interactions between hidden and visible sectors which generate $\mu$ and $B$ terms of comparable size once supersymmetry is broken. For simplicity, let us assume that SUSY is broken  by an $F$-term of a gauge singlet hidden sector field $X$. The most general \kahler potential allowed by symmetries is then 
\begin{equation}
 \Lagr_{B\mu}=\int d^4\theta \left(a \frac{X^\dagger}{\Mpl}  H_uH_d + b \frac{X^\dagger X}{\Mpl^2}H_u H_d +\hc\right)\,.
 \label{eq:SUGRAmu}
\end{equation}
Note that while these terms must respect a global symmetry imposed to forbid a large $\mu$, the symmetry is broken by the $F$-term of $X$. It is easy to see that (\ref{eq:SUGRAmu}) generates $\mu$ and $B$ given by
\begin{equation}
\label{eq:SUGRAmuscaling}
 \mu\sim a\frac{F^\dagger}{\Mpl},~~~B\sim b \frac{|F|^2}{\Mpl^2}\,.
\end{equation}
So far our discussion is very similar to the argument at the end of section \ref{sec:HiggsSector}. However, 
in the gravity mediation coupling constants $a$ and $b$ are both naturally of the order one. Combining this with an observation that (\ref{eq:SUGRAsoftterms}) and (\ref{eq:SUGRAmuscaling}) depend on the same dimensionful parameters, we conclude that the $\mu$-problem is solved in this model.

\subsection{Gauge Mediation}
\label{sec:gmsb}

{\it Minimal gauge mediation}
\vskip 0.2cm
To avoid the possibility that the Planck scale physics leads to observable flavor violation one could postulate that SUSY is broken the low energies and SUGRA contributions to the soft terms are negligible. To communicate SUSY breaking to the Standard Model fields, one then needs to introduce non-gravitational interactions between the hidden and visible sectors. If the two sectors interact only through the Standard Model gauge interactions, FCNC problem does not arise. This mechanism \cite{Dine:1981za,Dine:1993yw,Dine:1994vc,Dine:1995ag} is known as gauge mediated supersymmetry breaking (GMSB). It is instructive to start with a bottom-up approach to gauge mediation. We need to introduce new multiplets charged under all the Standard Model gauge groups. To avoid existing experimental constraints these messenger fields must be sufficiently heavy (which means that they must come in vector-like representations).
To communicate SUSY breaking the spectrum of messenger multiplets should be non-supersymmetric. This is achieved by assuming that messengers  couple 
to the SUSY breaking sector directly, either through Yukawa couplings or hidden sector gauge interactions.   We will also choose messengers in complete representations of the $SU(5)$ gauge group. While this choice has the benefit of maintaining successful gauge coupling unification, it is not strictly required and the $SU(5)$ language that we will use in the rest of the discussion is largely a convenient book-keeping device.
The simplest messenger content will contain $N$ flavors of messengers $Q$ and $\bar Q$ in $5$ and $\bar 5$ representations of $SU(5)$. The simplest way to parameterize messenger interactions with the SUSY breaking sector is by introducing a coupling to the SUSY-breaking spurion $X=M+\theta^2 F$:
\begin{equation}
\label{eq:messW}
 W_{mess}=XQ\bar Q\,.
\end{equation}
This form of the messenger spectrum is not the most general one, and the reader should consult the literature for examples of many interesting non-minimal models.
To construct a complete GMSB model with dynamical supersymmetry breaking in the hidden sector one usually promotes the spurion $X$ to a dynamical gauge singlet superfield.  One then introduces interactions between $X$ and the fields in the DSB sector that generate $X$ and $F_X$ vevs.

\begin{figure}
 \begin{center}
 \includegraphics[width=5in]{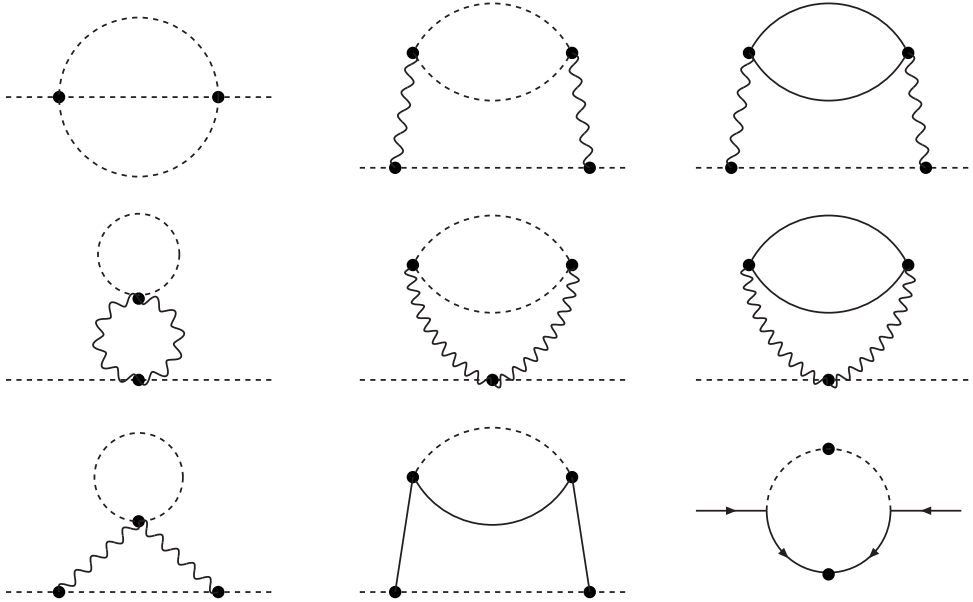}
 \caption{Diagrams giving rise to superpartner masses in gauge mediation.}
 \label{figgmsbsoft}
 \end{center}
 \end{figure}

The assumption that the messengers only interact with the Standard Model fields through gauge interactions implies that holomorphic soft terms, {\em i.e.} $A$-terms and $B$-term are parametrically small in GMSB models.
On the other hand, the Standard Model gauge interactions generate superpartner masses through processes shown in Figure \ref{figgmsbsoft}.  The resulting masses are given by \cite{Martin:1996zb}:
\begin{equation}
\label{eq:gmsbsoft}
\begin{aligned}
 &M_a=\frac{\alpha_a}{4\pi}N \frac{F}{M}g(x)\,,\\
 &\tilde m^2=2\left|\frac{F}{M}\right|^2\sum_a \left(\frac{\alpha_a}{4\pi}\right)^2 C_a N f(x)\,,
\end{aligned}
\end{equation}
where $a=1,2,3$ for $SU(3)$, $SU(2)$, and $U(1)_Y$ respectively, $C_a$ is a quadratic Casimir of a relevant scalar, $x=F/M^2$, and
\begin{equation} 
\begin{aligned}
 &g(x)&=&\frac{1}{x^2}\left((1+x)\log(1+x)+(1-x)\log(1-x)\right)\,,\\
&f(x)&=&\frac{1+x}{x^2}\left(\log(1+x)-2\mathrm{Li}_2\left(\frac{x}{1+x}\right)+\frac{1}{2}\mathrm{Li}_2\left(\frac{2x}{1+x}\right)\right)+(x\rightarrow -x)\,.
\end{aligned}
\end{equation}

It is often sufficient and convenient to work  in the limit of small SUSY breaking splitting within a messenger multiplet, $F\ll M^2$
\begin{equation}
\label{eq:mgmsbsmallF}
\begin{aligned}
& \tilde m^2=\sum_a \left(\frac{\alpha_a}{4\pi}\right)^a C_a N \left|\frac{F}{M}\right|^2\,,\\
 &M_a=\frac{\alpha_a}{4\pi} N \frac{F}{M}\,.
\end{aligned}\end{equation}

From (\ref{eq:gmsbsoft}) we see that sfermion and gaugino masses are generated at the same order in gauge couplings. Furthermore, requiring that the superpartner masses are at the electroweak scale we obtain a relation between the parameters of the messenger sector 
\begin{equation}\
\label{eq:FoverM}
 F/M \sim 100\, \TeV\,. 
 \end{equation}
On the other hand, gauge mediation allows a large range for the fundamental SUSY breaking scale,  $100\, \TeV<F_\text{DSB}<10^{10}\,\GeV$. The lower bound arises from the requirement that messenger mass squareds are positive and therefore $F>M^2$. Combining this with (\ref{eq:FoverM}) we conclude that the lower bound on both the messenger mass and the splitting within the messenger multiplet is of the order $100\, \TeV$. The requirement that the SUGRA contributions to soft masses are small compared to GMSB masses imposes an upper bound on the fundamental scale of SUSY breaking in the hidden sector, $F_\text{DSB}<10^{10}\,\GeV$.
Note that even if the messenger masses are near the lower bound, fundamental scale of SUSY breaking may be significantly higher: if messengers couple to the DSB sector weakly it is quite possible that $F\ll F_\text{DSB}$.
The discussion of SUSY breaking scale allows us to determine expression for gravitino masses in GMSB models
\begin{equation}
m_{3/2}=\left(\frac{\sqrt{F_\text{DSB}}}{100\,\TeV}\right)^2 2.4\mathrm{eV}\,.
\end{equation}

We chose messenger fields in the complete representations of the $SU(5)$ group to preserve one of the attractive features of the MSSM, gauge coupling unification. Insisting that the unification is perturbativity imposes an additional requirement --- if the messengers are light (with masses of \Order{100\TeV}) the number of messengers is restricted to be no more than five to avoid the Landau pole below the GUT scale.

\vskip 0.2cm
{\it Direct gauge mediation}
\vskip 0.2cm

Generating the necessary spectrum for messenger fields is  non-trivial. 
As we mentioned earlier, this can be achieved by promoting the spurion $X$ to a dynamical field and introducing interactions of $X$ with the DSB sector. Models of this type are often very complicated.
Another interesting approach involves attempts to construct models of direct gauge mediation where messengers  themselves play an essential role in SUSY-breaking dynamics. Realistic models of this type can be constructed if the DSB sector possesses a large global symmetry. Then one gauges an $SU(3)\times SU(2)\times U(1)$ subgroup of the flavor symmetry  and identifies it with the MSSM. 
Several viable examples of direct gauge mediation exist in the literature \cite{Poppitz:1996fw,Dine:2006xt,Kitano:2006xg,Csaki:2006wi}.

Let us illustrate direct gauge mediation with an explicit example \cite{Csaki:2006wi}. This model takes an advantage of the recent discovery of metastable SUSY breaking. As a DSB sector we will choose a massive SUSY QCD with $\tilde N$ colors and $F$ flavors in the magnetic description. The global $SU(F)$ symmetry of the theory is broken in a non-supersymmetric vacuum to an $SU(F-\tilde N)$ subgroup. We will embed the Standard Model gauge group into the unbroken global symmetry of the DSB sector. One can easily see that the contribution of the DSB sector to the Standard Model $\beta$-functions is $F$. To avoid Landau poles as long as possible we choose minumal values for $F=6$ and $\tilde N=1$. With this choice, the infrared physics of the DSB sector is described by an s-confined QCD rather than magnetic gauge theory. The electric gauge group, $SU(N)_{DSB}$,  has $N=F-\tilde N=5$ colors. Let us write down the matter content of the model in the magnetic description:
\begin{equation}
\begin{array}{|c|cccccccc|}
 \hline
&\phi&\bar\phi&\psi&\bar\psi&M&N&\bar N& X\\
\hline
SU(5)_\text{SM}&\fund&\afund&1&1&\bf Adj+1&\fund&\afund&1\\
\hline
\end{array}
\end{equation}
where $SU(5)_\text{SM}$ is an unbroken subgroup of the global $SU(6)$ symmetry.We will identify its $SU(3)\times SU(2)\times U(1)$ subgroup with the Standard Model.
We choose the superpotential
\begin{equation}
 W=\bar \phi M \phi +\bar\psi X\psi + \bar\phi N\psi+\bar\psi\bar N\phi-f^2 \Tr(M+X)\,.
\end{equation}

Clearly this model has the same matter content and superpotential as the one given in (\ref{eq:oraffnplusone}) with the identifications:
\begin{equation}
\begin{aligned}
 &B\rightarrow (\phi,\psi)\,,\\
 &\bar B \rightarrow (\bar \phi,\psi)\,,\\
&M\rightarrow \left(\begin{aligned}
&X&N\\
&\bar N&M
\end{aligned}
\right)\,.
\end{aligned}
\end{equation}
We conclude that supersymmetry is broken. DSB sector fields, $\phi$, $\bar\phi$, $N$, $\bar N$ and $M$,
are responsible for SUSY breaking but, once $SU(3)\times SU(2)\times U(1)$ subgroup of the global symmetry is weakly gauged, they also serve as messengers. However, the model is not fully realistic at this stage. 
This is due to the fact that the Coleman-Weinberg potential in this model leads to the ground state at $M=0$ and an  unbroken  accidental $R$-symmetry.
On the other hand,  $R$-symmetry breaking is required to generate gaugino masses (it is also required to generate masses for fermions in $M$ multiplet).

We need to modify the model so that $M$ acquires vev in the ground state and breaks the $R$-symmetry. This can be achieved by introducing new fields $S$, $\bar S$, $Z$, and $\bar Z$ with interactions\footnote{This superpotential breaks the $R$-symmetry explicitly.}
\begin{equation}
W=(d \Tr M+m)\bar S S+m^\prime(\bar Z S+S\bar Z)\,.
\end{equation}
Since new Coleman-Weinberg contributions to the potential of $\Tr M$ favor the minimum at $\Tr M=-m/d$ it is possible to choose the parameters of the Lagrangian so that the minimum is shifted away from the origin.
Another elegant way of breaking R-symmetry through gauge interactions was proposed in \cite{Dine:2006xt}.
To make the SUSY breaking in this model fully dynamical, additional dynamics must be introduced to  generate all mass terms in the superpotential \cite{Csaki:2006wi,Dine:2006gm}.
 
Finally, we should discuss gauge coupling evolution in this model. Formally gauge couplings unify since the messengers come in complete GUT representations. However, the effective number of messengers below the confinement scale of the DSB sector is 7. Above the confinement scale the effective number of messengers is 5. It is therefore clear that the QCD coupling will hit the Landau pole at a scale of the order of $10^{11}-10^{12}\GeV$.

One of the goals in GMSB model building is finding theories with a very low SUSY breaking scale.
In the model we just described this happens both by design and out of necessity --- unless SUSY breaking scale is low, gaugino masses are too small. The model predicts new light particles which could potentially be observable at future colliders.
On the other hand, the model is quite complicated and, most importantly, it can not be valid up to the GUT scale. These problems are general and often arise in other direct gauge mediation models based on metastable SUSY breaking that have been constructed recently \cite{Dine:2006xt,Kitano:2006xg}. 

\vskip 0.2cm
{\it General gauge mediation}
\vskip 0.2cm
While the microscopic physics describing various GMSB models may be quite different, their low energy phenomenology is usually very similar and can be described in terms of two parameters, an effective SUSY breaking scale, $\Lambda=F/M$ and an effective number of messengers $N$. As we saw, models of direct low energy gauge mediation introduce new interesting features --- additional light particles which, if we are lucky,  may be accessible at colliders.
More recently, it was realized \cite{Meade:2008wd} that GMSB phenomenology may be a lot richer. We will only briefly discuss these results  here. 
Following \cite{Meade:2008wd} we will define general gauge mediation (GGMSB) as a class of models where SUSY breaking sector decouples from MSSM in the limit of vanishing MSSM gauge couplings, $\alpha_a\rightarrow 0$.
This definition includes models of minimal and direct gauge mediation discussed earlier but it also includes strongly interacting theories. In such models, the perturbative calculations of superpartner masses are not reliable since the messengers themselves are strongly coupled. The authors of \cite{Meade:2008wd} analyzed GMSB contribution in terms of the correlation functions of gauge supercurrents and reached several important conclusions
\begin{itemize}
 \item The description of the most general gauge mediation model requires three complex parameters describing gaugino masses and three real parameters describing contributions to sfermion masses from each of the Standard Model gauge groups\footnote{In addition there is a possibility for D-term contribution to sfermion masses proportional to their hypercharge quantum numbers. However, such a contribution is dangerous since it generically leads to tachyonic slepton masses. It can be forbidden, for example, by invoking messenger parity \cite{Dimopoulos:1996ig}.}.
 \item Gaugino masses formally arise at tree level while sfermion masses squared arise at one loop. As a result it may be possible to construct feasible models with a fundamental scale of supersymmetry breaking as low as $10\,\TeV$. Unfortunately, such models are necessarily strongly coupled and calculation of the GGMSB parameters from the microscopic theory is not currently viable.
On the other hand, weakly coupled theories will generally have an additional suppression resulting in the usual scaling of superpartner masses with the Standard Model gauge couplings but the SUSY breaking scale must be at least $100\,\TeV$.
\item As a consequence of the new scaling of superpartner masses with the Standard Model gauge couplings, there may exist a hierarchy between gaugino and sfermion masses. Therefore, the definition of general gauge mediation encompasses gaugino mediated supersymmetry breaking \cite{Kaplan:1999ac}.
Moreover, in existing gaugino mediation models, supersymmetry is broken at a relatively high scale, and the renromalization group evolution leads to comparable sfermion and gaugino masses at the EWSB scale. On the other hand, since general GMSB models may break supersymmetry at very low energies, they could lead to a ``true'' gaugino mediated spectrum.
\end{itemize}

\vskip 0.2cm
{\it $\mu$ problem in gauge mediation}
\vskip 0.2cm
To illustrate the nature of the $\mu$-problem in GMSB we can review the argument at the end of section \ref{sec:HiggsSector} and identify the superfield $X$ in (\ref{eq:muX}) with the spurion that generates messenger mass. This means that in typical GMSB models (that is in the models with $F/X\sim 100\,\TeV$) the coupling constant $\lambda$ may be at most of the order $1/(16\pi^2)$. As we have seen such a small $\lambda$ implies unacceptably large $B$-term. There are several viable examples where $\mu$ and $B$ terms of the right size are generated without significant fine-tuning \cite{Dine:1995ag,Dvali:1996cu}. Several new ideas have been proposed recently \cite{Delgado:2007rz}. However, one can not say that a fully satisfactory solution for the  $\mu$-problem
in gauge mediation exists.

\subsection{Anomaly Mediation}
While the assumption of low scale supersymmetry breaking is attractive, it is not the only mechanism which may suppress flavor changing neutral currents. Even with the gravitino mass of the order TeV or larger, FCNCs may be suppressed if the Lagrangian has a sequestered form
\begin{equation}
\begin{aligned}
 &K=-3\Mpl^2\ln\left(1-\frac{f_\text{vis}}{3\Mpl^2}-\frac{f_\text{hid}}{3\Mpl^2}\right)\,,\\
 &W=W_\text{vis}+W_\text{hid}+W_0\,,\\
\end{aligned}
\end{equation}
where $f_\text{vis}$ and $f_\text{hid}$ are real functions of hidden and visible sector superfields respectively. Indeed, this form of the Lagrangian leads to vanishing of the soft terms at tree level. However, as we will see shortly, gaugino masses and $A$-terms are generated at one loop while the scalar mass squared are generated at two loops. This approach  \cite{Randall:1998uk} to communication of SUSY breaking is referred to as the anomaly mediated supersymmetry breaking (AMSB).

The sequestered form of the Lagrangian may be achieved in one of two ways.
The first approach \cite{Randall:1998uk} is based on the following  assumptions:
\begin{itemize}
 \item The fundamental theory lives in a 5-dimensional spacetime with one direction compactified on $S^1/Z_2$.
\item The hidden and visible sector fields are localized on different boundaries of the extra dimension.
\item There are no light bulk fields except for the fields in the supergravity multiplet.
\end{itemize}
With these assumptions, the locality of the low energy effective field theory guarantees the sequestered form of the effective Lagrangian.
While originally the 5D construction was suggested in the context of the flat 5D backgrouns, it may also be implemented within the Randal-Sundrum scenario. The AdS/CFT correspondence then suggests that there should exist a 4-dimensional realization of the theory. Such a realization was found in \cite{Luty:2001jh}. The hidden sector is assumed to be nearly conformal. One can then treat interactions between the hidden and visible sectors in (\ref{eq:mediationkahler}) as small perturbations of strong conformal dynamics in the hidden sector. As the hidden sector approaches the infrared fixed point, coupling constants, $c_{ff^\prime}$, become negligibly small as a consequence of RG flow.

Given the sequestered form of the Lagrangian, one can integrate out dynamics of the hidden sector and parameterize supersymmetry breaking by an $F$-term of an auxiliary superfield in the supergravity multiplet, referred to as a compensator superfield\footnote{It is conventional to work in units of $\Mpl$ and the $F_\Phi$ has an unusual dimension one, in fact $F_\Phi=m_{3/2}$.}
\begin{equation}
 \Phi=1+F_\Phi \theta^2\,.
\end{equation}

The compensator superfield couples to the MSSM fields according to
\begin{equation}
 \Lagr=\int d^4\theta \Phi^\dagger \Phi K(Q^\dagger, e^V Q)+\left(\int d^2\theta \Phi^3W(Q)+\hc \right)\,.
\end{equation}
We can see the effects of sequestering here --- the visible sector Lagrangian appears completely supersymmetric if we perform  holomorphic field redefinition to write it in terms of a rescaled superfield $\tilde Q= \Phi Q$. This result is a consequence of scale invariance of MSSM Lagrangian at the tree level. However, at the quantum level scale invariance is lost. To maintain formal scale invariance we need to rescale not only light fields but also the cutoff scale of the theory, $\Lambda_{UV} \rightarrow \Phi\Lambda_{UV}$. This last rescaling results in the appearance of soft terms in the visible sector.

The most straightforward way to derive soft masses relies on a method of analytic continuation to the superspace \cite{ArkaniHamed:1998kj}. Let us begin with gaugino masses. Writing down gauge kinetic terms with rescaled cutoffs and expanding the gauge coupling function in powers of $\theta^2$ we obtain
\begin{equation}
 \int d^2\theta \frac{1}{4g_i^2(\mu/\Lambda_\text{UV}\Phi)} \Wg^\alpha \Wg_\alpha 
=\int d^2\theta \left(\frac{1}{4g^2(\mu/\Lambda_\text{UV})} - \frac{b_i}{32\pi^2}\ln \Phi\right) \Wg^\alpha \Wg_\alpha\,,
\end{equation}
where $b_i$ is a one loop $\beta$-function coefficient. Expanding the log in the last term and performing superspace integral we obtain gaugino mass  \cite{Randall:1998uk,Giudice:1998xp}
\begin{equation}
 m_{\lambda_i}(\mu)=\frac{b_i}{2\pi}\alpha_i(\mu) F_\Phi\,.
\end{equation}

Soft scalar masses are obtained by starting with an expression for the renormalized \kahler potential  \cite{Randall:1998uk}
 \begin{equation}
  \int d^4\theta Z\left(\frac{\mu}{\Lambda_\text{UV}(\Phi^\dagger \Phi)^{1/2}}\right)Q^\dagger Q\,.
 \end{equation}
Expanding in powers of $\theta^2$ leads to 
\begin{equation}
\label{eq:amsbsfermion}
 \tilde m^2_f(\mu) =-\frac{1}{4}\frac{\pd\gamma_f(\mu)}{\pd \ln \mu}|F_\Phi|^2=
\frac{1}{4}\left(\frac{b_i}{2\pi}\alpha^2_i\frac{\pd\gamma_f}{\pd \alpha_i}+\frac{b_\lambda}{2\pi}\frac{\lambda^2}{(4\pi)^2}\frac{\pd \gamma_f}{\pd \alpha_\lambda}\right)\,|F_\Phi|^2
\end{equation}
where
\begin{equation}
 \gamma_f(\mu)=\frac{\pd \ln Z(\mu)}{\pd \ln \mu}
\end{equation}
is the anomalous dimension of the sfermion, $\lambda$ is the Yukawa coupling, $b_\lambda$ and $b_i$ are one loop coefficients of gauge and Yukawa couplings respectively, and one needs to sum over all gauge and Yukawa couplings of the field in question.
One can similarly obtain trilinear scalar soft terms  \cite{Randall:1998uk}
\begin{equation}
\begin{aligned}
 A^u_{ff^\prime}=\frac{1}{2}\left(\gamma_u(\mu)+\gamma_f(\mu)+\gamma_{f^\prime}(\mu)\right)\lambda^u_{ff^\prime}F_\Phi\,,\\
 A^d_{ff^\prime}=\frac{1}{2}\left(\gamma_d(\mu)+\gamma_f(\mu)+\gamma_{f^\prime}(\mu)\right)\lambda^d_{ff^\prime}F_\Phi\,,
\end{aligned}
\end{equation}
where $\gamma_u(\mu)$ and $\gamma_d(\mu)$ are anomalous dimensions of $H_u$ and $H_d$ respectively.

As we can easily see, AMSB is an extremely predictive theory --- the soft parameters are given in terms of $F_\Phi$ and the Standard Model gauge and Yukawa couplings at the TeV scale. The minimal model is insensitive to UV physics. Let us illustrate this by adding\footnote{To slightly simplify the argument we will assume that new fields do not couple to MSSM in the superpotential.} to MSSM a set of heavy fields , a vector-like fourth generation with mass $m_H$:
\begin{equation}
 \Lagr_H=\int d^4\theta \Phi^\dagger \Phi K(Q_H^\dagger, e^V Q_H) +\left(\int d^2\theta \Phi^3 m_H Q_H\bar Q_H+\hc\right)\,.
\end{equation}
We have included coupling to the compensator field $\Phi$.
Naively, it appears that new fields might affect AMSB predictions since they modify $\beta$-functions at high scales $\mu\gg m_H$.
 However, the tree level Lagrangian of heavy superfields depends on $\Phi$ even after holomorphic rescaling --- the spectrum of the heavy supermultiplet is not supersymmetric. In fact, identifying $m_H \Phi$ with the spurion $X$ of GMSB models, we see that heavy superfields play a role of messengers.
The soft masses in the infrared are given by the sum of the high energy AMSB contribution and gauge mediated contribution of the new fields. It is easy to check that to the leading order in $F_\Phi$ the soft parameters in the IR, $\mu\ll m_H$,  are completely determined by $\beta$-functions and coupling constants of the low energy theory.
 
Unfortunately, the minimal AMSB model can be immediately ruled out. Slepton masses squared given by (\ref{eq:amsbsfermion}) are negative! It turns out to be extremely difficult to modify AMSB models to fix the slepton mass problem --- the difficulty is due to celebrated UV-insensitivity of anomaly mediation. A number of solutions \cite{Pomarol:1999ie} to this problem were proposed over the years. However, while many of these solutions are viable, none of them seem sufficiently compelling as they are typically quite complicated and almost necessarily sacrifice the UV-insensitivity of the anomaly mediation.

\section{Acknowledgements}
I would like to thank Tao Han and K. T. Mahantappa, as well as students and lecturers at TASI 2008 for providing great and stimulating atmosphere at. This work was supported in part by NSF grant No. PHY-0653656.